\documentclass[12pt,preprint]{aastex}
\newcommand{\noprint}[1]{}

\shorttitle{Multisite observations of pulsating sdB stars}
\shortauthors{Reed et al.}

\begin{document}

\title{Follow-up observations of pulsating subdwarf B stars:
Multisite campaigns on PG~1618+563B and PG~0048+091}

\author{M. D. Reed \altaffilmark{1}, S. J. O'Toole\altaffilmark{2,3}, 
D.M. Terndrup\altaffilmark{4}, J. R. Eggen\altaffilmark{1}, 
A.-Y. Zhou\altaffilmark{1}, D. An\altaffilmark{4}, C.-W. Chen\altaffilmark{5},
W. P. Chen\altaffilmark{5}, H.-C. Lin\altaffilmark{5},
C. Akan\altaffilmark{6}, O. Cakirli\altaffilmark{6},
H. Worters\altaffilmark{7,8}, D. Kilkenny\altaffilmark{7}, 
M. Siwak\altaffilmark{9},
S. Zola\altaffilmark{9,10},
Seung-Lee Kim\altaffilmark{11}, G. A. Gelven\altaffilmark{1},
S. L. Harms\altaffilmark{1}, and G. W. Wolf\altaffilmark{1}}
\affil{$^1$ Department of Physics, Astronomy, \& Materials Science, Missouri
State University, 901 S. National, Springfield, MO 65897 U.S.A.}
\email{MikeReed@missouristate.edu}
\affil{$^2$ Anglo-Australian Observatory, PO Box 296, Epping NSW 1710, Australia}
\affil{$^3$ Dr Remeis-Sternwarte, Astronomisches Institut der Universit\"at
  Erlangen-N\"urnberg, Sternwartstr. 7, Bamberg 96049, Germany}
\affil{$^4$ The Ohio State University, 140 W. 18th Avenue, Columbus, OH
43210, U.S.A.}
\affil{$^5$ Graduate Institute of Astronomy, National Central University, 
Chung-Li, Taiwan}
\affil{$^6$ Ege University Observatory, 35100 Bornova-Izmir, Turkey}
\affil{$^7$ South African Astronomical Observatory, PO Box 9, Observatory 7935, 
South Africa}
\affil{$^8$ Centre for Astrophysics, University of Central Lancashire,
Preston, PR1 2HE, UK}
\affil{$^{9}$ Astronomical Observatory, Jagiellonian 
University, ul. Orla 171, 30-244 Cracow, Poland}
\affil{$^{10}$ Mt. Suhora Observatory of the Pedagogical University, ul.
Podchor\c{a}zych 2, PL-30-084 Cracow, Poland}
\affil{$^{11}$ Korea Astronomy and Space Science Institute, Daejeon, 305-348, 
South Korea}

\begin{abstract}
We present follow-up observations of pulsating subdwarf B (sdB) 
stars as part of our
efforts to resolve the pulsation spectra for use in asteroseismological
analyses. This paper reports on multisite campaigns of the pulsating
sdB stars  PG~1618+563B and PG~0048+091.
Data were obtained from observatories placed around the globe for
coverage from all longitudes. For PG~1618+563B, our five-site campaign 
uncovered a dichotomy of pulsation states: Early during the campaign the
amplitudes and phases (and perhaps frequencies) were quite variable while
data obtained late in the campaign were able to fully resolve five stable
pulsation frequencies. For 
PG~0048+091, our five-site campaign uncovered a plethora of frequencies
with short pulsation lifetimes. We find them to have observed properties
consistent with
stochastically excited oscillations, an unexpected result for subdwarf B
stars.  We discuss our findings and their impact
on subdwarf B asteroseismology.
\end{abstract}

\section{Introduction}
Subdwarf B (sdB) stars are thought to have masses 
about 0.5M$_{\odot}$, with thin ($<$$10^{-2}$M$_{\odot}$) hydrogen shells 
and temperatures from $22\,000$ to $40\,000$~K (Saffer et al. 1994).
They are horizontal branch stars that have shed nearly all of
their H-rich outer envelopes near the tip of the red giant branch and 
as He-flash
survivors, it is hoped that asteroseismology can place constraints on 
several interesting phenomena.
Subdwarf B star pulsations come in two varieties: short period (90 to 600
seconds; EC~14026 stars after that prototype, officially V361~Hya stars, or
sdBV stars) with
amplitudes typically near 1\%, and long period (45 minutes to 2 hours;
PG~1716 stars after that prototype or LPsdBV stars)
with amplitudes typically $<$0.1\%. For more
on pulsating sdB stars, see Kilkenny (2001) and Green et al. (2003) for
observational reviews and Charpinet, Fontaine, \& Brassard (2001) for a 
review of
the proposed pulsation mechanism.
For this work, our interest is the sdBV
(EC~14026) class of pulsators.

In order for asteroseismology 
to discern the internal conditions of variable stars, the pulsation ``mode''
must be identified from the temporal spectrum (also called pulsation spectrum or Fourier
transform; FT). The mode is represented mathematically by spherical harmonics 
with quantum numbers $n$ (or $k$), 
$\ell$, and $m$. For nonradial, multimode pulsators the
periods, frequencies, and/or the spacings between them are
most often used to discern the
spherical harmonics (see for example Winget 1991). 
These known modes are then
matched to models that are additionally constrained by
non-asteroseismic observations, typically $T_{\rm eff}$
and $\log\,g$ from spectroscopy. Within such constraints, the model that
most closely reproduces the observed pulsation periods (or period spacing)
for the constrained modes
is inferred to be the correct one. Occasionally such models can be
confirmed by independent measurements (Reed et al. 2004,
Reed, Kawaler, \& O'Brien 2000, Kawaler 1999), but
usually it is impossible to uniquely identify
the spherical harmonics and asteroseismology cannot be applied to obtain a
unique conclusion. Such has been the case for sdBV stars, which
seldom show multiplet structure (i.e., even frequency spacings) that may be used 
to observationally 
constrain the pulsation modes. However, relatively few sdBV stars have been
observed sufficiently to know the details of their pulsation spectra. The
goal of our work is to fully resolve the pulsation spectrum, search for
multiplet structure, and examine the 
characteristics of the pulsation frequencies over the course of our
observations.

In this paper we report on multi-site follow-up
observations of the pulsating sdB stars
PG~1618+563B (hereafter PG~1618B) and PG~0048+091 (hereafter PG~0048) obtained
during 2005.
PG~1618B was discovered to be a variable star by Silvotti et al. (2000;
hereafter S00)
who  detected frequencies of 6.95 and 7.18 mHz 
($P\approx 144$ and 139 s respectively) from short data runs ($<2.3$~hrs)
obtained during
seven nights, three of which were separated by three months. 
PG~1618 is an optical double consisting of a
main sequence F-type star (component $A$) with an sdB star (component $B$)
at a separation of 3.7 arcseconds. The combined brightness is $V=11.8$ while the
sdB component has $V\approx 13.5$. The discovery
data used a combination of photoelectric photometry,
which did not resolve the double, and CCD data which did. The 
combined flux of the double 
in the former would have reduced the pulsation amplitudes. From spectra
obtained at Calar Alto, S00 determined that $T_{\rm eff}=33\,900\pm
1\,500K$ and $\log g=5.80\,\pm 0.2$.

PG~0048 was discovered to be a variable star during 10 observing runs
varying in length from 1 to 4 hours obtained in 1997
and 1998 (Koen et al. 2004; hereafter K04). As they used a variety of
instruments spread over the span of a year, K04 were only able to 
combine two consecutive runs, from which they detected seven frequencies.
However, it is obvious from their temporal spectrum 
that pulsation amplitudes and possibly frequencies
were changing (their Fig. 3). K04 contributed this to unresolved frequencies
caused by their short duration data runs.
K04 also obtained an optical spectrum and examined 2MASS colors to
determine that PG~0048 has a G0V-G2V companion; though the orbital
parameters are unknown and no
$T_{\rm eff}$ or $\log g$ estimates were given.

Here we report the results of a new program to resolve the pulsation spectra of these
two stars. Section~2 describes the observations,
reductions, and  analysis for PG~1618B, and \S 3 the same for PG~0048. 
Section 4 compares the results for both stars and discusses the
implications for asteroseismology.

\section{PG~1618B+563B}
\subsection{Observations}
PG~1618B was observed from 5 observatories 
(Baker, MDM, McDonald, Lulin, and Suhora) over a 45 day period during
spring 2005.
Data obtained at MDM
(2.4~m) and McDonald (2.1~m) observatories used the same 
Apogee Alta U47+ CCD camera.
This camera is connected via USB2.0 for high-speed readout, and our
binned ($2\times 2$) images had an average dead-time of one second.
Observations at Baker (0.4~m) and Lulin (1.0~m) observatories were 
obtained with Princeton Instruments RS1340 CCD cameras. Data obtained at 
Baker Observatory were                    
binned $2\times 2$ with an average dead-time of one second, while 
observations from  Lulin Observatory used a $392\times 436$ subframe at
$1\times 1$ binning with an average dead-time of six seconds. The 
Mt. Suhora Astronomical Observatory (0.6~m) data were obtained 
with a photomultiplier tube photometer which has microsecond dead-times. 
Observations obtained at McDonald, Baker, and MDM observatories used a
red cut-off (BG40) filter, so the transmission is virtually the same as
the blue photoelectric observations from Suhora observatory. Observations from
Lulin Observatory used a Johnson $B$ filter, which slightly reduced the
amount of light collected compared to other observations, but does not
impose any significant phase and/or amplitude changes compared to other
observations (Koen 1998; Zhou et al. 2006). 
Accurate time was 
kept using NTP (Baker, McDonald, and MDM observatories) or GPS
receivers (Lulin and Suhora observatories) and corrected to barycentric time during
data reductions.

Standard procedures of image reduction, including bias subtraction,
dark current and flat field correction, were followed using
{\sc iraf}\footnote{ {\sc iraf} is distributed by the National Optical Astronomy
Observatories, which are operated by the Association of Universities
for Research in Astronomy, Inc., under cooperative agreement with the
National Science Foundation.} packages.
Differential magnitudes were extracted from the calibrated
images using {\sc momf} (Kjeldsen \& Frandsen 1992) or occasionally they
were extracted using {\sc iraf} aperture photometry with extinction
and cloud corrections using the normalized intensities of several field
stars, depending on conditions. Photoelectric data reductions proceeded using standard
Whole Earth Telescope reduction packages (Nather et al. 1990).
As sdB stars are substantially hotter, and thus bluer, than typical field
stars, differential light curves using an ensemble of comparison stars
are not flat due to differential atmospheric and color extinctions.
A low-order polynomial was fit to remove these trends from the data on a
night-by-night basis. Finally, the lightcurves are normalized by their
average flux and centered around zero so the reported differential
intensities are $\Delta I=\left( I/\langle I \rangle\right)-1$.
Amplitudes are given as milli-modulation amplitudes (mma) with an
amplitude of 10~mma corresponding to an intensity change of 
1.0\% or 9.2~millimagnitudes.

The companion of PG~1618B
adds a complication to the reductions in that data obtained at McDonald and MDM
observatories resolved the optical double, but those from other observatories did not. 
Using our data for which the stars are resolved,
we determined that component A contributes 67.3\% of the total flux. To correct
the unresolved data, we created a fitting function by smoothing the data
over many points (around 50 points per box), multiplying it by 0.673 and
subtracting it from the unresolved data. While this process effectively
removes the flux from PG~1618A, it cannot correct for the noise of this
component, which remains behind. As such, the corrected data are
noisier, limiting their usefulness.

\begin{figure}
\figurenum{1}
\includegraphics[angle=-90,width=\textwidth]{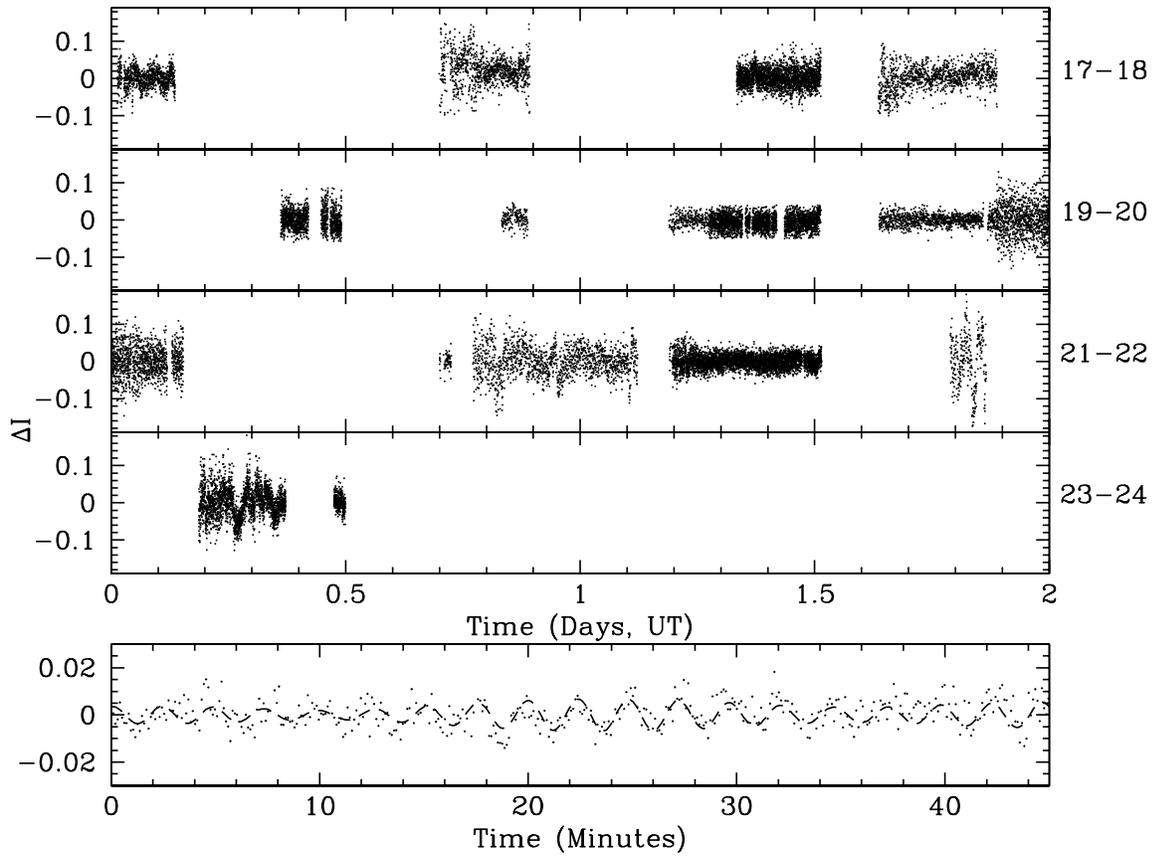}
\caption{Lightcurves for PG~1618B data. The top four panels show the coverage
during the multisite portion of the campaign from March 17 -- 24, 2005,
while the bottom panel shows
an enlarged section of a run obtained at MDM. Note that the scales are
different for the bottom panel. The line indicates our fit to the data.}
\label{fig01}
\end{figure}

Multiple-longitude coverage was only obtained during the
first week of the campaign. A total of
73.5 hours of data were collected from three observatories (McDonald, 
Lulin, and Suhora) which provided a
47\% duty cycle. Subsequent data were obtained only in Missouri
(Baker Observatory) and Arizona (MDM Observatory). These data serve to extend 
the timebase of observations (increasing the temporal resolution) and to decrease
the noise in the temporal spectrum. Lightcurves showing the coverage of 
the first six nights of observations, as well as a portion of a typical
MDM run, are provided in Fig.~\ref{fig01}.

\placefigure{fig01}

\placetable{tab01}

\subsection{Analysis}
\begin{figure}
\figurenum{2}
\includegraphics[angle=-90,width=\textwidth]{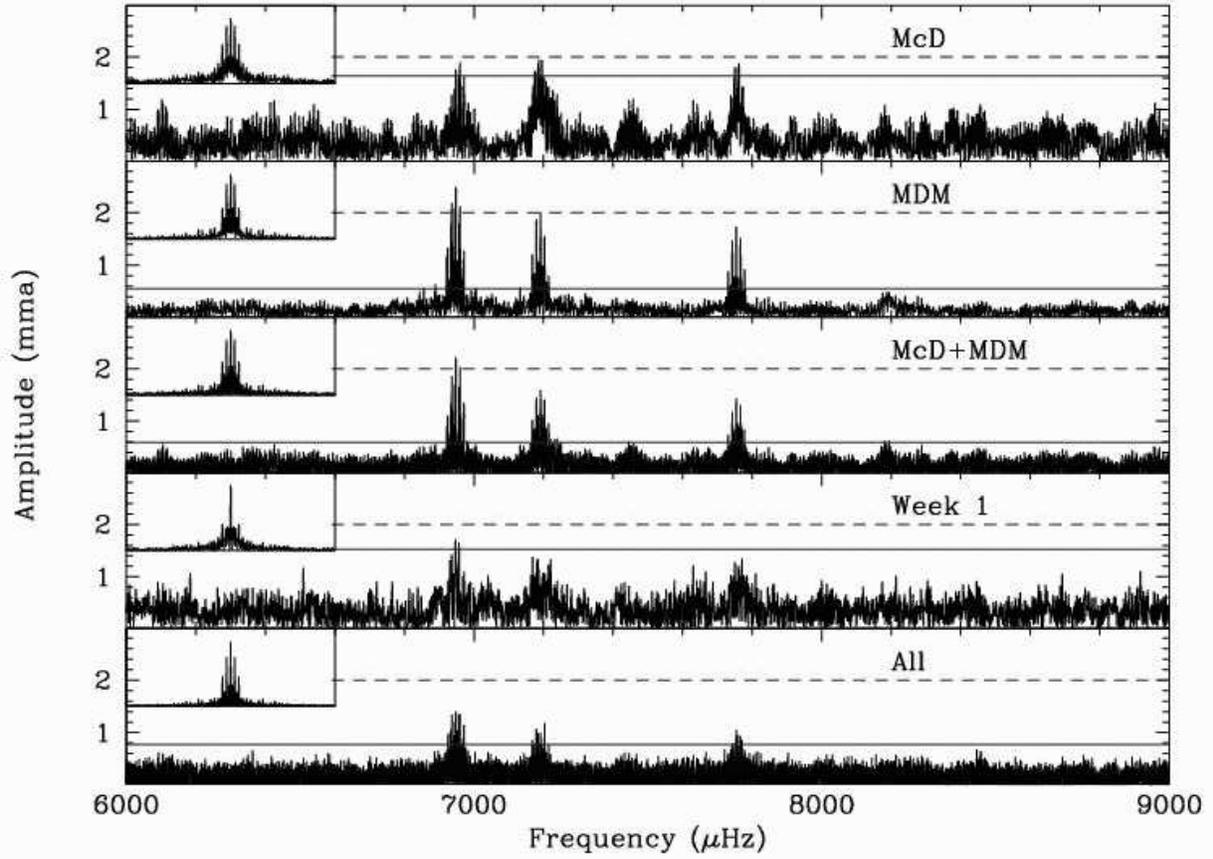}
\caption{Temporal spectra of various subsets of PG~1618B data. Insets
are the window functions. Solid (blue) horizontal line is the $4\sigma$
detection limit while the dashed (blue) lines are at 2~mma in all
panels.}
\label{fig02}
\end{figure}

Our campaign was quite long (about 45 days) with a concentration of
data at the beginning, but the best data (highest S/N and best conditions)
were obtained at the end. We therefore
grouped combinations of nightly runs into the subsets given in
Table~\ref{tab02} for analysis. Table~\ref{tab02} also provides the
temporal resolution (calculated as 1/$t_{\rm run}$ where $t_{\rm run}$
is the length of the observing run) and the $4\,\sigma$ detection limit 
(calculated using areas adjacent to the pulsation but outside of their
window functions). The
temporal spectra  and window functions of these subsets
 are plotted in Fig.~\ref{fig02}. A window function is a
single sine wave of arbitrary, but constant amplitude
sampled at the same times as the data. The central peak of the window
is the input frequency, with other peaks indicating the
aliasing pattern of the data. Each peak of the data spectrum 
intrinsic to the star will create such an aliasing pattern. As is 
evident from Fig.~\ref{fig02}, the MDM data was significantly better than
the rest, so we began our analysis with that subset.

\placefigure{fig02}

\placetable{tab02}

\begin{figure}
\figurenum{3}
\includegraphics[angle=-90,width=\textwidth]{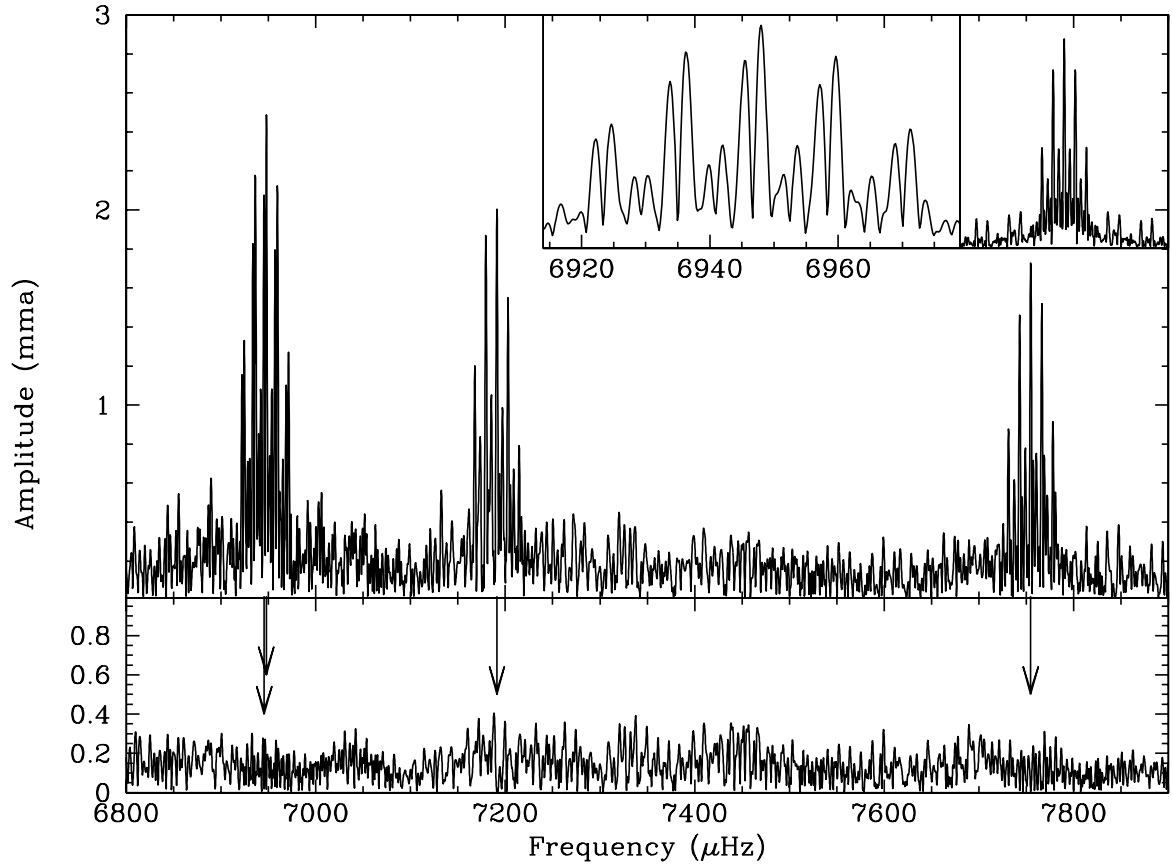}
\caption{Temporal spectrum of PG~1618B data. Top panel shows the original FT
of the combined MDM data while the bottom panel shows the residuals after
prewhitening. Prewhitened frequencies are indicated by arrows. Insets show an
enlarged view of the frequency doublet and the window function.}
\label{fig03}
\end{figure}

Analysis of the MDM data was relatively easy and straightforward.
In Fig.~\ref{fig03}, the top panel shows the
original FT, while the bottom panel shows the residuals after prewhitening
by the frequencies indicated by arrows.
The insets show the window function (top right) and an expanded view of
a 65$\mu$Hz region around the close doublet.
 Frequencies,
amplitudes and phases were determined by simultaneously fitting a nonlinear
least-squares  
solution to the data. Since during the MDM observations the amplitudes
were relatively constant, the solution proceeded as expected and prewhitening
effectively removed the peaks and their aliases. The formal 
solution and errors for the MDM subset are given in Table~\ref{tab03}.

\placefigure{fig03}

\placetable{tab03}

Examination of the other data sets 
indicates that while PG~1618B was well-behaved during the MDM
observations, it was not at other times. This was most noticeable in
our examination the data collected during the first week. While the 
temporal spectrum has the cleanest window function, \emph{no} peaks
are detected above the $4\sigma$ detection limit (1.53~mma)
even though peaks are detected in individual Lulin and McDonald runs 
 (see Fig.~4 which will be discussed 
in \S 2.3). Combining the well-behaved MDM data with
any other data set results in a decrease of amplitudes, indicating
that outside of the MDM data, the amplitudes, phases, or even
frequencies are not constant. If the pulsation properties were consistent
throughout the campaign, data collected at smaller telescopes,
with low S/N would still have been useful for reducing the overall noise. 
Unfortunately, such was not the case so
all we can really conclude is that the MDM data
detected all the pulsations that were occurring (above the
detection limit) at that time while the pulsations intrinsic to PG~1618B
\emph{must} have been more complex at other times. It would
be interesting to study the longer-term variability of
PG~1618B, but using only 2~m-class telescopes.

Outside of the combined data sets, there are two frequencies that are
detected above the $4\sigma$ detection limit during individual runs. The least-squares 
solutions for these frequencies are provided at the end of Table~\ref{tab03}.
The frequency at $\approx 9199\,\mu$Hz was above the noise only in the
March 22 McDonald data, though a peak at the same frequency also appears in 
the Suhora data during March 16 and 21. The frequency at $\approx 8179\,\mu$Hz 
was above the noise only in the April 30 MDM data though corresponding
peaks appear in the McDonald March 18, Lulin March 18, and Suhora March 21
runs. Since they are detected above the $4\sigma$ detection criteria for those
runs, we include them in our discussion that follows.

\subsection{Discussion}
Silvotti et al. (2000) detected two frequencies in their discovery data while
we clearly resolve four frequencies from our MDM dataset, and
two more from individual data runs, bringing the total to six independent
frequencies.
We calculate the S00 resolution to be 5.5~$\mu$Hz and 
estimate their noise to be about 1~mma though this is
misleading in that because of their short data runs, their window function
effectively covers all of the remaining pulsations. However, for a strict
comparison, we can say that our MDM data alone are $3\times$ better
in resolution and have a detection limit twice as good, though in a practical
sense our MDM data are far superior solely based on the duration of our
individual data runs. Had S00 observed for longer durations (particularly
with their CCD setup), their data would likely have been similar to the same
number of runs from our MDM set. However, it is clearly safe to say that
our MDM data alone are insufficient to describe the complexity of 
pulsations occurring within PG~1618B. As such, the remainder of our
discussion which is based on the MDM data, can only be a \emph{minimum}
of what is really occurring.

\subsubsection{Constraints on the pulsation modes}
One of our goals is to observationally identify or constrain the
pulsation modes of individual frequencies. 
Differing $m$ components
of the same degree $\ell$ have degenerate frequencies unless perturbed,
typically by rotation. If a star is rotating, then each
degree will separate into a multiplet of $2\ell +1$ components with
spacings nearly that of the rotation frequency of the star.
As such, observations of multiplet structure can constrain the pulsation
degree (for examples, see Winget et al. 1991 for pulsating white
dwarfs and Reed et al. 2004 for sdB stars). For PG~1618B, there are no
two frequency spacings that are similar, though there are not many
frequencies to work with. The lack of observable multiplet structure
is typical of sdBV stars but is likely limited to
four possibilities: i) Rotation is sufficiently
slow that all $m$ values remain degenerate within the frequency resolution of our
data; ii) our line of sight is along the
pulsation axis, with $\sin i\approx 0$, leaving only the $m=0$ mode
observable because of geometric cancellation (Pesnell 1985; Reed, Brondel,
\& Kawaler 2005); iii) rapid internal rotation is such that $m$ multiplets are
widely spaced and uneven (Kawaler \& Hostler 2005); or iv) at most one
pair is part of a multiplet with an unobserved component of the multiplet.

Spectroscopy can only rule out large splittings for possibility (i) as
spectroscopic limits are typically $\approx 10$~km/s and from Fig.~1 of
S00, PG~1618B appears as a ``normal'' sdB devoid of rapid rotation.
Possibility (ii) can only be determined for cases in which the sdB star is
part of a close binary such that the rotation and orbital axes can be
inferred to be aligned. Since PG~1618 is only an 
optical double at wide separation,
it does not constraint the alignment of the surface spherical
harmonics.  Similarly, possibility (iii) is virtually impossible to decipher
unless the star pulsates in many (tens of) frequencies, and would still require
some interrelation of spacings for modes of the same degree (Kawaler \& 
Hostler 2005). Possibility
(iv) also remains an option, though a difficult one to constrain.
Higher resolution (and perhaps longer duration)
spectroscopy would help to answer this question, and multicolor photometry
or time-series spectroscopy might also be able to discern the spherical
harmonics (see Koen 1998 and O'Toole et al. 2002 for examples of each).

Another quantity that can be used to constrain the pulsation modes is  
the frequency density. Using the assumptions that no two frequencies share
the same $n$ and $\ell$ values (except possibly the close pair at
$6946\,\mu$Hz), and that high-degree $\ell\geq 3$ modes are not observationally
favored because of geometric cancellation (Charpinet et al. 2005; Reed, 
Brondel, \& Kawaler 2005), we can ascertain whether the frequencies are too
dense to be accounted for using only $\ell\leq 2$ modes. From stellar models, a general
rule of thumb is to allow three frequencies per $1000\,\mu$Hz. We will ignore
$f1$ which is too distant in frequency space and count $f5$ and $f6$
as a single degree $\ell$. This leaves four frequencies within 1235~$\mu$Hz;
which can easily be accounted for using only $\ell\leq 2$ modes. Indeed,
even if $f5$ and $f6$ do not share their $n$ and $\ell$ values, the
frequency spectrum can still accommodate all of the detected frequencies
without invoking higher degree modes. Of course this does not mean
that they are not $\ell\geq 3$ modes, only that the pulsation spectrum
is not sufficiently dense to require their postulation.

\subsubsection{Amplitude and phase stability}
If pulsating sdB
stars are observed over an extended time period,
it is common to detect amplitude variability in many, if not all, of the
pulsation frequencies (eg. O'Toole et al. 2002; Reed et al. 2004; Zhou et al. 2006).
Such variability can occasionally be ascribed to beating between
pulsations too
closely spaced to be resolved in any subset of the data. However, variations often
appear in clearly resolved pulsation spectra where
mode beating cannot be the cause. For PG~1618B, frequencies $f1$ and $f2$ are only
detected during a single run each and frequencies $f5$ and $f6$ are too
closely spaced to be resolved during individual runs, leaving only 
frequencies $f3$ and $f4$ available for analysis of amplitude variations.

\begin{figure}
\figurenum{4}
\includegraphics[angle=-90,width=\textwidth]{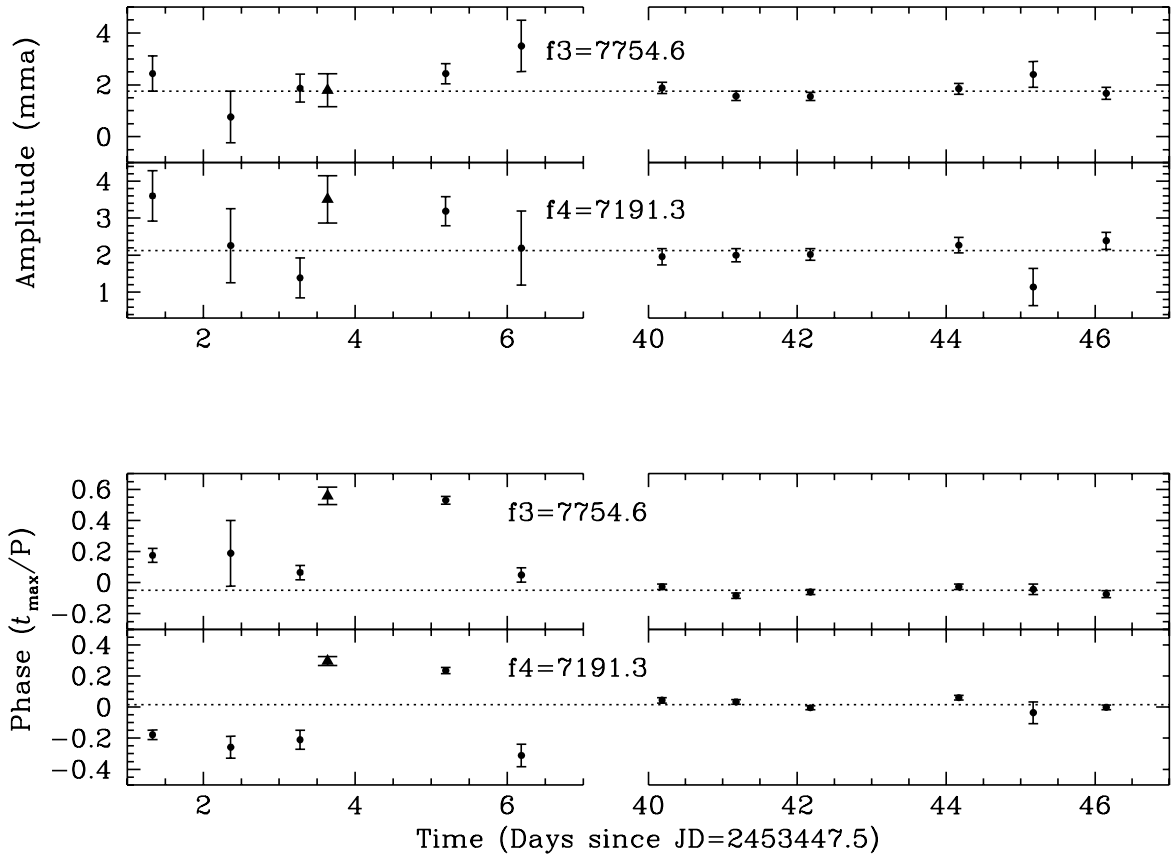}
\caption{Amplitudes and phases of the two frequencies in PG~1618B that are
resolvable nightly. The data are only those from McDonald and MDM
observatories except for a single Lulin run (marked by a triangle). 
The dashed lines indicate our fits to the MDM data.
Note that the time axis is discontinuous.}
\label{fig04}
\end{figure}

Figure~\ref{fig04} shows the amplitude and phases of these two frequencies
for individual data runs from
McDonald and MDM observatories as well as a single Lulin run (marked by a
triangle); these frequencies were not detected elsewhere. 
During the MDM observations, the amplitudes and phases
for both frequencies are nearly constant (to within the errors)
 except for one low amplitude, but they have a significant variation
in the McDonald and Lulin data. Of particular interest are the phases and
amplitudes of $f4$, especially those during day three, in which we
have both a McDonald and Lulin run that do not overlap in
time. Between these two runs, the amplitude, which had been decreasing
during the previous three days, suddenly increases to begin the same
declining pattern again. The phases also show a bimodal structure
early in the campaign with phases near $-0.20$ and $+0.25$ with the
first phase jump occurring coincident with the amplitude
increase. Except for the lack of sinusoidal amplitude variation, this
has the appearance of unresolved pulsations. However, if the two
unresolved frequencies had intrinsic amplitude variability, then it
\emph{could} reproduce the observations. However the MDM observations,
which are not only steady, but have $f4$ phases intermediate to the
McDonald and Lulin data, do not support this. Clearly, during the week
of MDM observations, PG~1618B had neither amplitude nor phase variations
and since the MDM phases do not coincide with phases from earlier in
the campaign, unresolved pulsations are unlikely. Since the data obtained
at MDM and McDonald observatories used the same acquisition system and
time server (NTP), errors in timing also seem unlikely.

\placefigure{fig04}

\section{PG~0048+091}
\subsection{Observations}
We originally observed PG~0048 as a secondary target during 
a campaign on KPD~2109+4401
(Zhou et al. 2006). Those data revealed a complex pulsation 
spectrum which we could
not resolve with such limited sampling and a short time base.
As such, PG~0048 was re-observed as a multisite campaign during Fall 2005.
Five observatories participated in the campaign with the specifics of 
each run provided in Table \ref{tab04}. Though we were a bit
unlucky with weather, over the course of 
our 16 night campaign we obtained 167.4 hours of data for a duty cycle of 44\%.
Details of the observing instruments and configurations are the same as
for PG~1618B,
except for the following: 
SAAO (1.9~m) used a frame transfer CCD with millisecond
dead-times but only an $\approx 30\times 40$ arcsecond field of view,
which resulted in no comparison stars within the CCD field. 
As such no transparency variations
could be corrected and only photometric nights were used.
Tubitak Observatory used a Fairchild CCD447 detector; during the
first run the images had $1\times 1$ binning with a dead-time of 102 seconds,
while subsequent runs used $2\times 2$ binning with a dead-time of
29 seconds. Bohyunsan Optical Astronomy Observatory (BOAO 1.9~m) data 
were obtained with a SiTe-424 CCD windowed to $580\times 445$ pixels, binned 
$2\times 2$ with an average dead time of 14 seconds. 
MDM and SAAO used red cut-off filters, making their
responses very similar to blue-sensitive photoelectric 
observations, while Lulin, BOAO, and
Tubitak used no filter making their sampling more to the red. 
As pulsations
from sdB stars have little amplitude dependence in the visual 
and no phase dependence (Koen 1998; Zhou et al. 2006),  
mixing these data is not seen as a problem.

\begin{figure}
\figurenum{5}
\includegraphics[angle=-90,width=\textwidth]{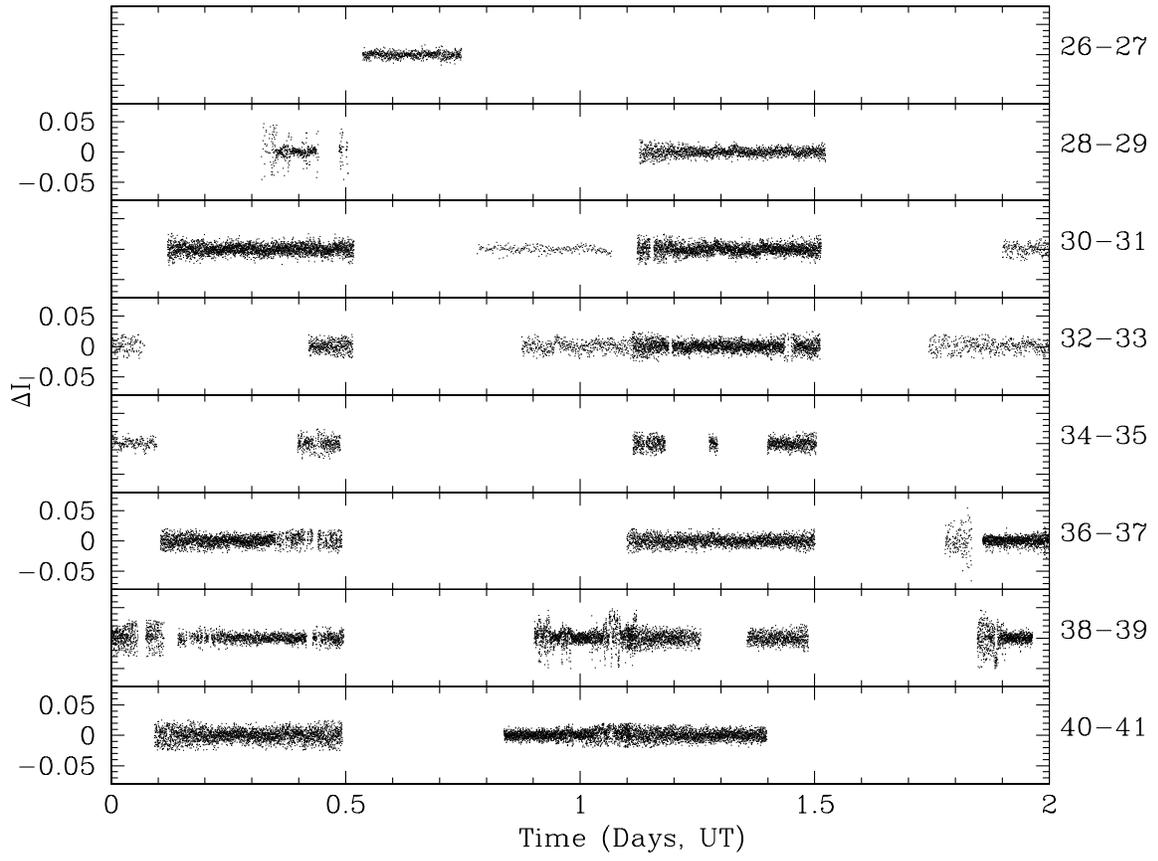}
\caption{Lightcurves for the PG~0048 data. Each panel is two days.}
\label{fig05}
\end{figure}

The standard procedures of image reduction, including bias subtraction,
dark current and flat field correction, were followed using {\sc iraf}.
Differential magnitudes were extracted from the calibrated
images using {\sc momf} (Kjeldsen \& Frandsen 1992), except for the SAAO
data for which we used aperture photometry because there were no comparison
stars.
As described for PG~1618B, we again used
low-order polynomials to remove airmass trends between our blue target star and
the redder comparison stars.
The lightcurves are normalized by their average
flux and centered around zero, so the reported differential intensities are
$\delta I = \left( I/\langle I\rangle \right) -1.$ 
Figure~\ref{fig05} shows the lightcurve of PG~0048 with each panel covering two days.

\placetable{tab04}

\placefigure{fig05}

\subsection{Analysis}
\begin{figure}
\figurenum{6}
\includegraphics[angle=-90,width=\textwidth]{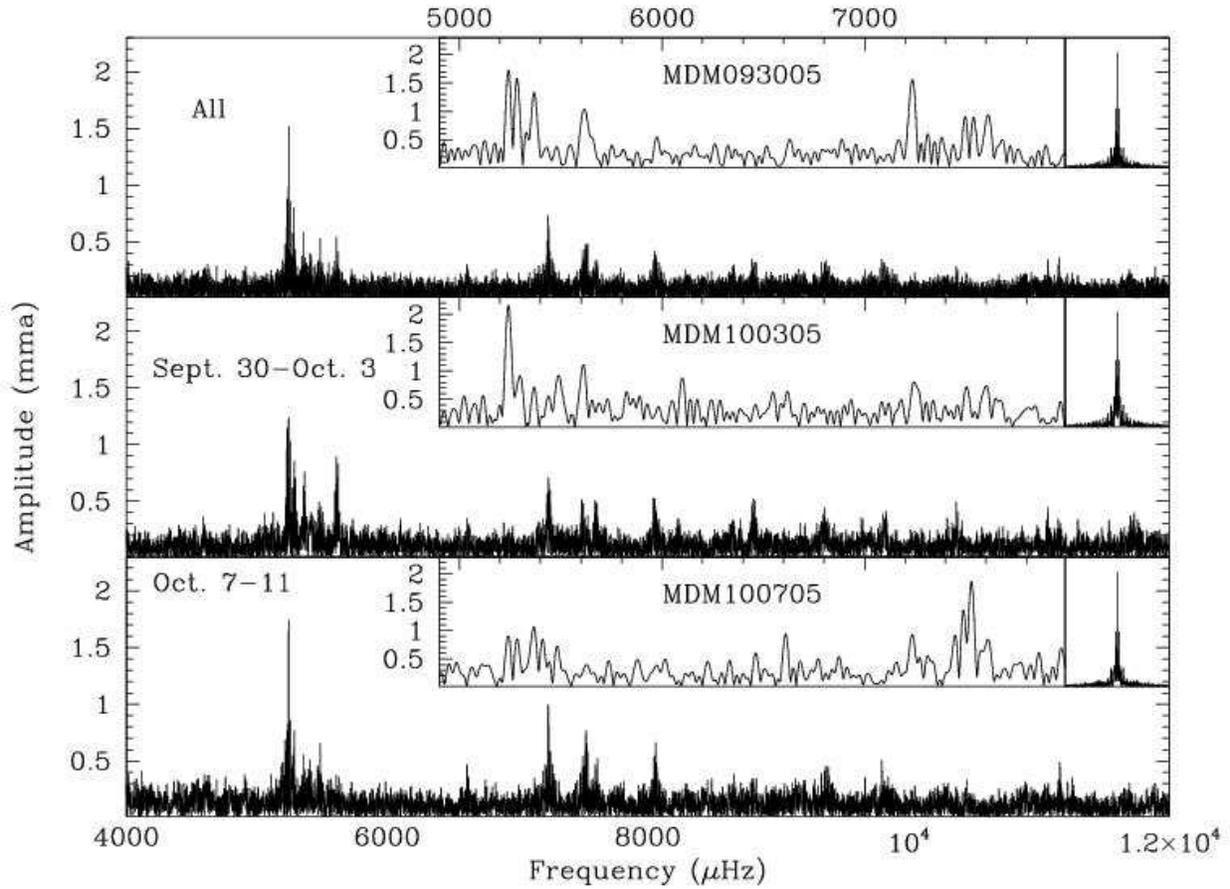}
\caption{Comparison of temporal spectra for varying groups of PG~0048 data. 
The large panels show FTs for combinations of data (inclusive dates are 
labeled). The insets on the right are 
the window functions for the combinations of data to scale. The central insets are
slightly enlarged FTs of individual runs obtained congruent with the larger panels and are
labeled. They go from early to late within the campaign.}
\label{fig06}
\end{figure}

During the campaign, we completed a ``quick-look'' analysis
of data runs as early as possible to ascertain the data quality and the
pulsation characteristics of the star. We noticed early on
that the temporal spectra of PG~0048 changed
on a nightly basis with pulsation frequencies appearing and then disappearing
on subsequent nights.  
 Likewise, we knew that our analysis would be complicated
by severe amplitude variations which would limit the usefulness of
 prewhitening techniques
and could create aliasing. Figure~\ref{fig06} shows the effects of amplitude
variations. The full panels are pulsation spectra for three groups of data:
All of the data; data obtained from September 30 through October 3; and
from October 7 through October 11. The right insets are the 
corresponding window functions plotted on the same horizontal 
scale. At such large scales, the windows appear as single peaks and
show that the changes in the FTs are not caused by aliasing. 
The central insets are
individual data runs within the larger set and show the
  variability between runs. When sets of data are combined in
which the peak amplitudes are not constant, an FT will 
show the average amplitude.
For the frequencies that appear in only a few runs, the amplitudes are
effectively quashed in the combined FT. As PG~0048 is the most pulsation 
variable sdB star currently known, our immediate goal is to glean as many 
observables from these data as possible. While we do provide some
interpretation, our aim is to provide sufficient information for theorists
to test their models.

\placefigure{fig06}

\begin{figure}
\figurenum{7}
\includegraphics[angle=-90,width=\textwidth]{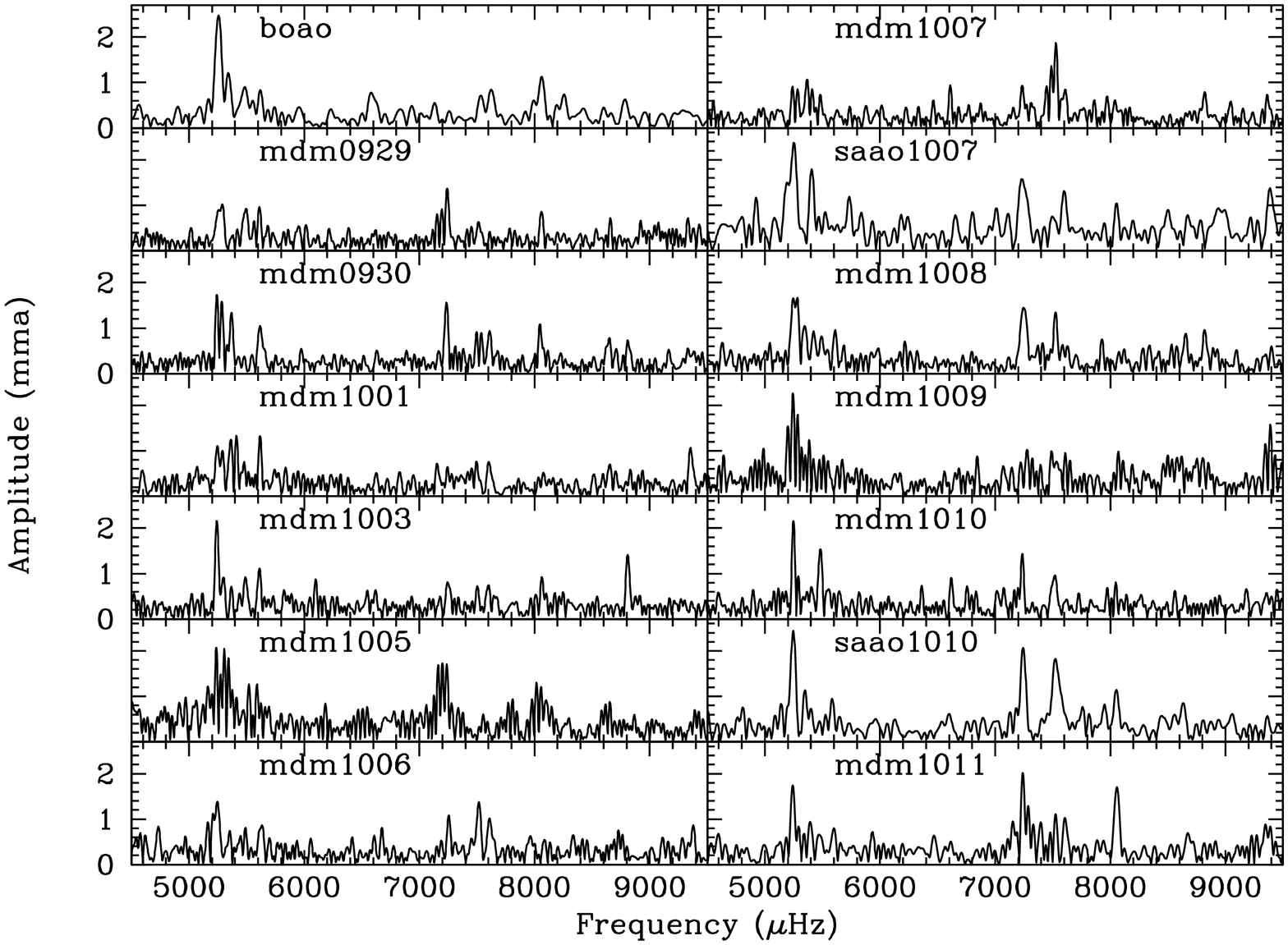}
\caption{Temporal spectra for individual PG~0048 observing runs.}
\label{fig07}
\end{figure}

The complexity of the data meant it was necessary to analyze it using
multiple techniques: We performed standard Fourier analyses on
combined sets of observations to increase temporal resolution and lower
the overall FT noise and analyzed
individual runs of the best quality data. The analysis of individual runs
represents a time-modified Fourier analysis, which is essentially a Gab\'or
transform, except that we replace a Gaussian time discriminator with the
natural beginnings and endings of the individual runs. As the best 
individual runs are not continuous with time (and nearly all are from MDM
Observatory) the use of a Gaussian-damped traveling temporal wave discriminator
(a standard Gab\'or transform) would not enhance the results. 
The temporal spectra of these
runs are shown in Fig.~\ref{fig07}. Runs mdm1005 and mdm1009, though long
in duration, have gaps in them because of inclement weather, whereas the other 
12 runs are gap-free. For these 12 runs
aliasing in the FT is not a problem and the only constraints are the 
width of the peaks, which are determined by run length, and the noise of
the FT, which is a combination of the signal-to-noise of each point and the
number of data points within the run.

\placefigure{fig07}

Frequencies, amplitudes and phases were determined using two different
software packages, Period04 (Lenz \& Breger 2004) and a custom (Whole
Earth Telescope) set of non-linear least squares fitting and
prewhitening routines. Each of the three data combinations in
Fig.~\ref{fig06}, the 12 gap-free 
data runs plotted in Fig.~\ref{fig07}, the three
data runs obtained during 2004, and the 10 runs from the discovery data
(kindly provided by Chris Koen) were reduced using both
software packages. Overall, more than 35 frequencies were fit during
at least one data run. Table~\ref{tab05} provides information for 28
frequencies which have been detected above the $4\sigma$ detection limit.
Column 1 lists a frequency designation; column 2 the frequency as fit
to the highest temporal resolution data set in which each frequency 
is detected with
the formal least-squares errors in column 3. Column 4 provides
the standard deviation of the corresponding frequencies detected in
individual runs and column 5 gives the number $N$ of 
individual runs in which that frequency was detected (from twelve 2005 runs 
and three 2004 runs).
Tables~\ref{tab06} and \ref{tab07} provide the corresponding amplitudes
as fit for individual runs and various combinations of data 
acquired during the  2005 campaign, a re-analysis of the discovery data,
 and the 2004 MDM data.
The last two rows of these tables
provide the $4\sigma$ detection limits and temporal resolutions
for the runs. Our
determination that these frequencies are real and 
intrinsic to the star is based on
i) detection by both fitting software packages, and amplitude(s)
higher than the $4\sigma$ detection limit with ii) detection
during several data runs, and/or iii) detection at amplitudes too large
to be associated with aliasing.

\placetable{tab05}

\placetable{tab06}

\placetable{tab07}

\subsection{Discussion}
\subsubsection{Frequency content}
\begin{figure}
\figurenum{8}
\includegraphics[angle=-90,width=\textwidth]{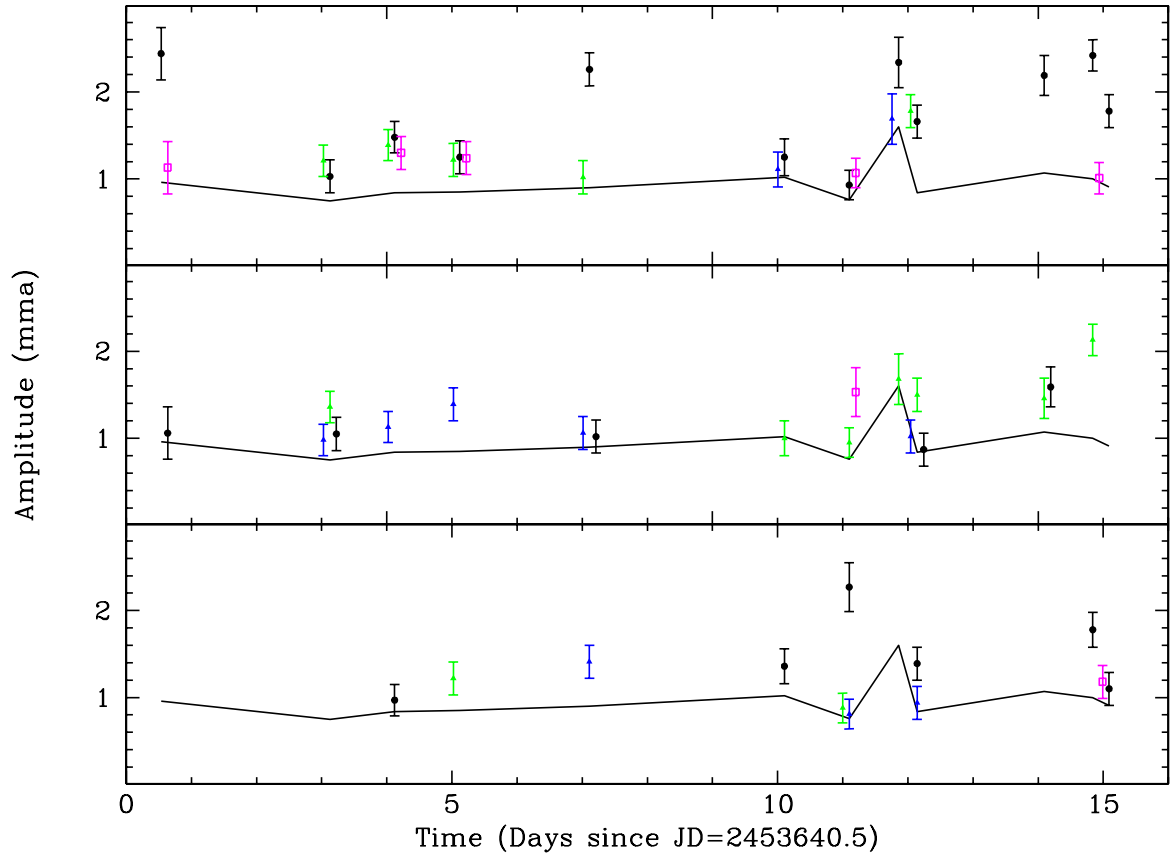}
\caption{Amplitudes and errors of 12 frequencies detected in PG~0048.
Each panel contains amplitudes for four frequencies (differentiated by point 
type and color) which may be shifted by $\pm 0.1$~day if they overlap. The 
solid line is the $4\sigma$ detection limit.}
\label{fig08}
\end{figure}

\begin{figure}
\figurenum{9}
\plottwo{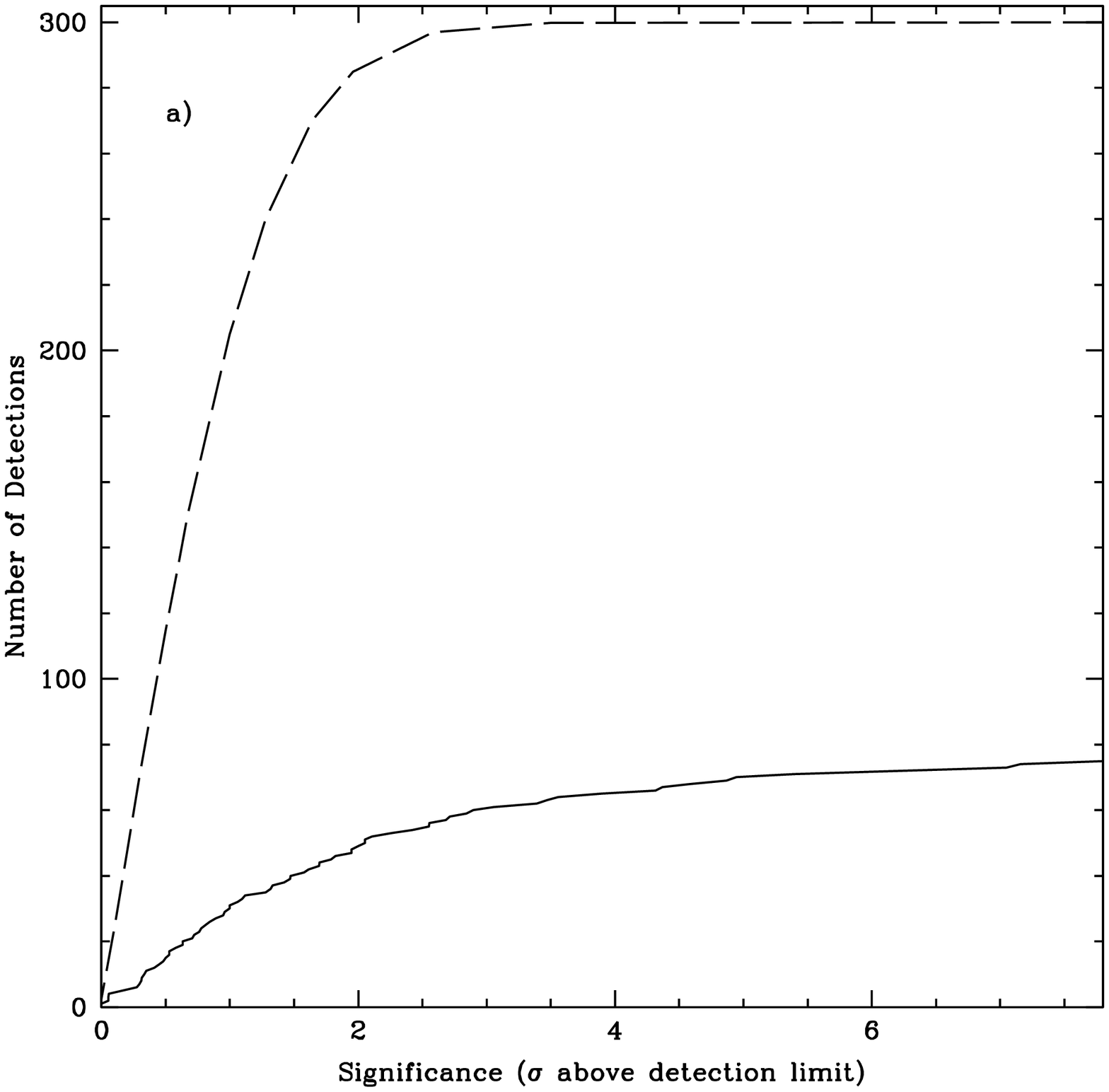}{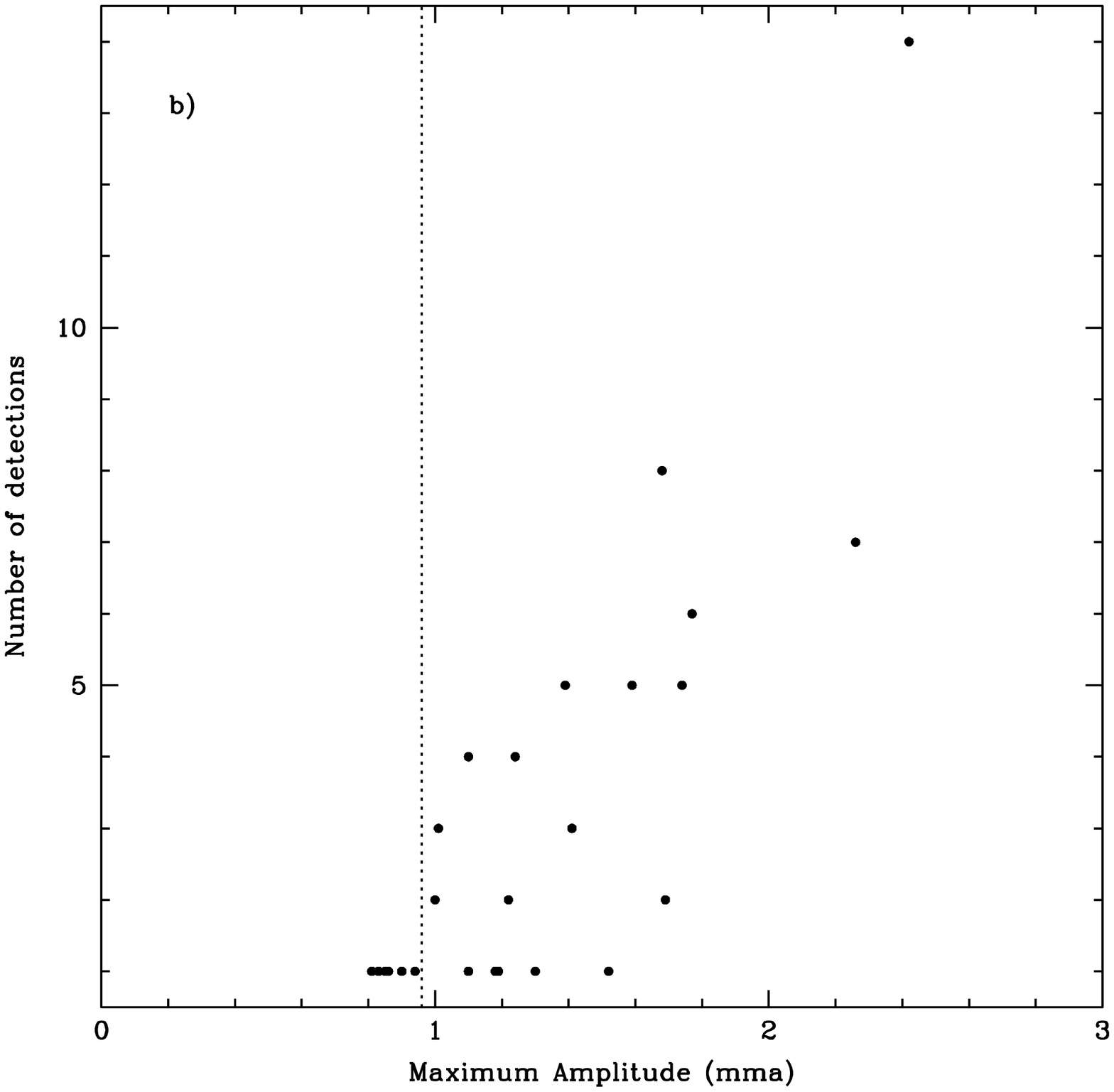}
\caption{a) The number of detections of frequencies from individual runs for
PG~0048 compared to Gaussian probability.
The number of possible detections (dashed line) compared to the amount
of actual detections (solid line) depending on the detection significance.
b) A comparison between the number of detections for individual frequencies
and the maximum amplitude detected.
\label{fig09}}
\end{figure}

During our 2005 multisite campaign, we detected 24 pulsation frequencies
from individual data runs plus an additional frequency from the
combined data set (which was also detected in 2004). We recover all
seven of the frequencies detected in the discovery data (K04), but
only 14 of the 16 frequencies detected from our 2004 data.
As can be seen from Table~\ref{tab05}, PG~0048 shows an atypically
large range of frequencies for sdBV-type  
pulsators, %
%\footnote{Excluding the two known pulsating sdB stars that have
%both short and long periods.}
%%%%%%%%% The above is not really relevant, since (a) the long periods
%%%%%%%%% may not be sdBV-like oscillations, and (b) the frequency
%%%%%%%%% ranges are actually still quite short compared to
%%%%%%%%% PG0048. BA0901 has lots of combo frequencies but otherwise
%%%%%%%%% the range is smaller than PG0048.
especially
when considering that only one frequency, $f_{23}$ (11103.3\,$\mu$Hz), can be
identified as a linear combination (of $f_6$ and $f_7$). 
%Additionally,
%it has a very rich and relatively distributed pulsation spectrum:
%except for regions around 5300 and 7300~$\mu$Hz, there are very few
%groups of frequencies. 
As noted in Table~\ref{tab05}, only one frequency is
detected in all of our data ($f_2$: $\sim$5244.9\,$\mu$Hz) while
the next most common frequency ($f_{13}$: $\sim$7237.0\,$\mu$Hz) is
detected in only 11 of the 15 runs. Several frequencies are
only detected once or twice (e.g. $f_{12}$: $\sim$7154.3\,$\mu$Hz), and 
so we should test if their amplitudes are
sufficient to consider them to be real. If a particular frequency has amplitudes
that are $1\sigma$ above the detection limit, 
then we might only expect to detect it
68\% of the time\footnote{This is a lower limit since statistically, the pulsation
amplitude is equally likely to be \emph{higher} than $1\sigma$ from the detected
level rather then below it.}. Figure~\ref{fig08} shows the amplitudes and $1\sigma$
errors for 12 different frequencies (four frequencies per panel) and the detection 
limit (solid line) for individual runs. The (black) circles in the top panel are 
for $f2$, which is detected in every run. However, the two frequencies
indicated by (magenta) squares in the middle and bottom panels are only detected
once, even though they are $>1\sigma$ above the detection limit. If their
amplitudes were nearly constant (to within their errors), they would be detected
at least 68\% of the time. Another way to show this is in panel $a$ of
Fig.~\ref{fig09} where
the detections are plotted against their significance. We detect a total of 24
frequencies from 12 individual runs from our 2005 data. If we detected all 24
frequencies from every run, we would have made 288 detections, while we only actually
made 75. The solid line shows the
number of individual detections cumulative with significance (the number of 
standard deviations the detection was above the detection limit). In other words,
49 of our 75 detections were $2\sigma$ or less above the detection limit. The
dashed line is the standard Gaussian probability distribution which shows that 
at $1\sigma$ significance, we should have made at least 196 (68\%) detections.
Since 64\% of our detections are $\geq 1\sigma$, our 75 detections is well short
of what we should have detected, indicating that the pulsation amplitudes are
\emph{really} falling below the detection limit. Panel $b$ compares the number
of actual detections to the maximum pulsation amplitude. As expected, there is
some correlation as the higher the amplitude, the easier it is to detect
that frequency.
Additionally, the highest amplitude (and therefore most easily detected)
frequency is the same during 1997, 1998, and 2004 ($f8$: 
$\sim$5612.2\,$\mu$Hz) but is only detected in $\sim 1/3$ of our 2005
data runs during which $f2$ ($\sim$5244.9\,$\mu$Hz) had the highest amplitudes. 
This change in pulsation amplitudes will be further discussed in \S 3.3.3.

\placefigure{fig08}

\placefigure{fig09}

\subsubsection{Constraints on mode identifications}
\begin{figure}
\figurenum{10}
\includegraphics[angle=-90,width=\textwidth]{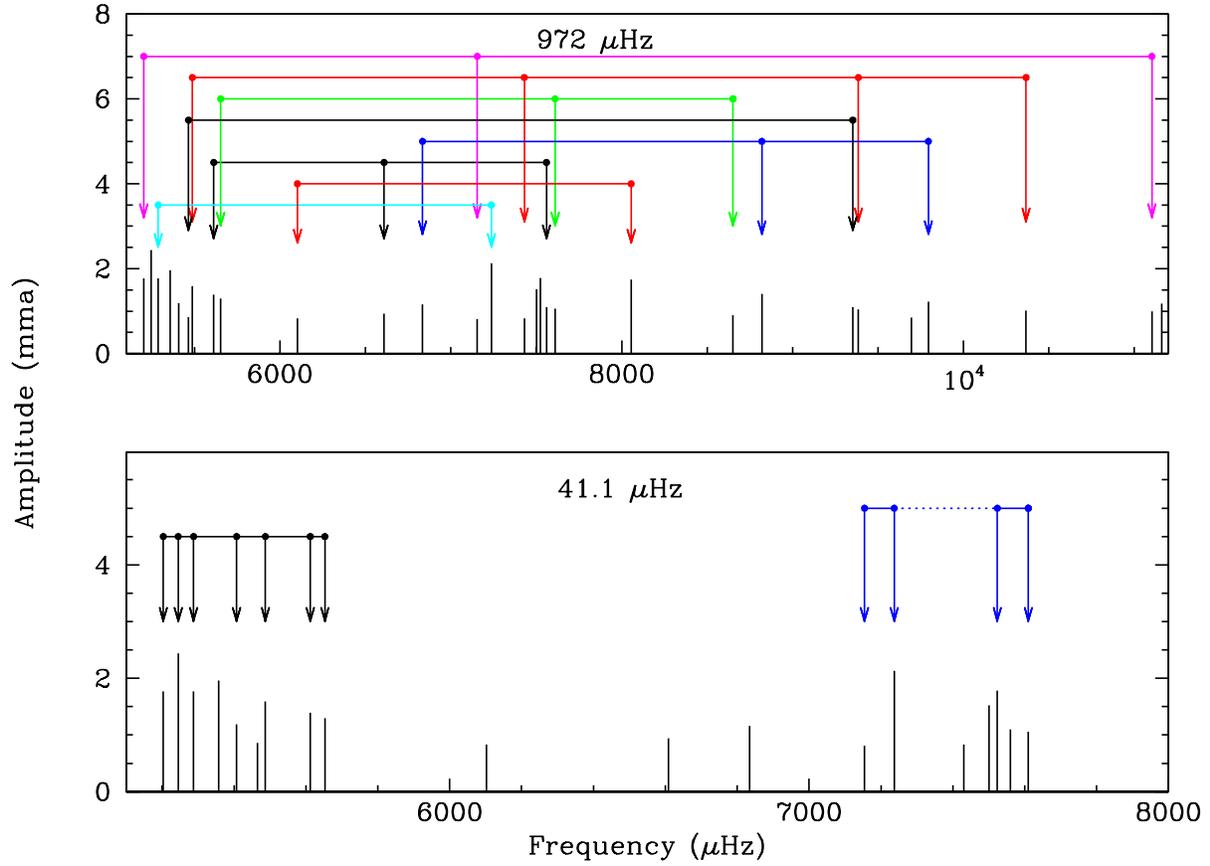}
\caption{A schematic of the pulsation content of PG~0048 to indicate
the nearly evenly spaced frequencies (provided in
Table~\ref{tab08}). The top panel shows the spacings that are
integer multiples near $972~\mu$Hz
and the bottom panel shows those near $41.1~\mu$Hz. For clarity,
each set of multiplets are connected via horizontal bars at differing 
heights and have different colors (electronic version
only, and black is used twice in the top panel).}
\label{fig10}
\end{figure}

As in \S 2.3.1, one of the best ways to relate pulsation frequencies to
pulsation modes is via multiplet structure. With such a rich pulsation
spectrum, it seems likely that some of the frequencies should be 
related by common frequency splittings.
If not, then the pulsation spectrum is too dense
(discussed below) for the frequencies to consist of 
only low-order $\ell \leq 2$ modes. 
There are two different spacings that occur
many times with splittings near 972 and 41.1~$\mu$Hz. 
Table~\ref{tab08} lists these 
frequencies and the deviation from the average spacing between them,
while Fig.~\ref{fig10} shows them graphically. While the spacings may be
important,
it is difficult to attach any physical meaning to them.
A spacing of $972~\mu$Hz is far too large to be associated with
stellar rotation, as it corresponds to a rotation period of only 17 minutes.
There is currently no high-resolution spectrum of PG~0048,
  and the star's main-sequence companion would complicate any attempts to
  measure its rotation velocity. However, the typical $v\sin i$ of
  sdBs is less than $\sim$\,5\,km\,s$^{-1}$ (Heber, Reid, \& Werner 2000). 
If we try to explain this
large splitting using asymptotic theory (consecutive $n$ overtones rather
than $m$ multiplets),
we would expect successive overtones of the radial index to be
  roughly evenly spaced for large $n$, with one series for each $\ell$
degree. But the spacings observed in
  PG~0048 are irregular, requiring \emph{eight} degrees (one for every
line in Table~\ref{tab08}), and $3\leq\ell\leq 7$ modes have reduced
visibility because of geometric cancellation (Charpinet et al. 2005; Reed,
Brondel, \& Kawaler 2005). Theory suggests that
sdBV stars should pulsate in low overtone modes and the spacings between 
successive low
  overtone modes should differ by several hundreds of $\mu$Hz
(Charpinet et al. 2002), which again does not match what is observed. 

\placetable{tab08}

\placefigure{fig10}

Another asymptotic-like relation 
is the ``Kawaler scheme'' which has been recently
presented by Kawaler et al. (2006) and Vu\u{c}kovi\'{c} et
al. (2006). Though it is
not compatible with the low overtone ($n$) pulsation theory associated
with sdB pulsators, an improved
frequency fit can sometimes be obtained using an asymptotic-like formula;
$$ f(i,j)=f_o+i\times\delta +j\times\Delta$$
where $i$ has integer values, $j$ is limited to values of $-1,\,0,$ and $1$,
$\delta$ is a large frequency spacing and $\Delta$ a small one. 
However, for the case of PG~0048, the small spacings are unrelated to
the larger spacings and the larger spacings themselves do not interrelate,
but rather appear in sets with differing spacings between the
sets. So this scheme is not applicable for PG~0048.
%This, and the large scatter in night-to-night
%  frequencies, means the Kawaler scheme cannot be applied to PG~0048.

%%%%%%% NEED TO INSERT CORRECT VUCKOVIC ET AL. (2006) REFERENCE

The smaller,
41.1~$\mu$Hz spacings are more akin to what we would expect for rotationally
split multiplets, though still large compared to typical measured rotation
rates. If the low-frequency set
are all components of a single multiplet, it would require very high
degree ($\ell\geq 5$) pulsations, which are not observationally
favored (Charpinet et al. 2005; Reed, Brondel, Kawaler 2005). The same
is true for the high-frequency set \emph{if} $f13$ and $f16$ belong
to the same set. Since they are separated by $7\times 41.0\mu$Hz, it
is certainly possible that $f12$-$f13$ and $f16$-$f18$ are just two
pairs, but if they are all combined into a single multiplet, it would
require $\ell\geq 5$ again. Yet these could also be just chance
superpositions.

% Since PG~0048 pulsates in so many frequencies, we felt it was important
% to test the significance of detection for the above spacings. A 5\% 
% error of a 972~$\mu$Hz splitting is a rather large 49~$\mu$Hz and
% 41.1~$\mu$Hz is near our resolution limit
% (29-30\,$\mu$Hz) for most of our individual
% data runs. To test their significance, we produced Monte Carlo simulations,
% randomly placing 28 frequencies within 6000~$\mu$Hz and counting
% how often 14 frequency splittings could be found within a 5\% range.
% We produced more than a million simulations which fit the observations
% 125\% of the time; that is, there were more than 14 such splittings produced
% from nearly \emph{every} simulation, mostly at large frequency splittings,
% with a bit over 30\% producing two such splittings per set of frequencies.
% However, our simulations were fairly simple and we estimate that $\sim$10\% of 
% the detections were close doublets matched to a single frequency and
% counted twice. Though crude, the simulations reveal that while
% a splitting near 972~$\mu$Hz occurs 14 times, and a splitting near
% 41.1~$\mu$Hz occurs 9 times, it seems very likely that these are just
% chance superpositions of frequencies caused by a rich pulsation
% spectrum and have only been included for completeness.
Since PG~0048 pulsates in so many frequencies, it is important to test the
significance of the spacings discussed above. We did this by producing Monte
Carlo simulations, randomly placing 28 frequencies within 6000 microHz of each
other, and counting how often we could detect 14 frequency splittings the
same to within about 5\%. This criterion would find all but one of the
splittings we actually observe. After analysing over one million simulations,
we detected \emph{at least} 14 splittings in nearly every case. In other
words, the splittings we observe are \emph{not} statstically significant.

Another tool we can use to place constraints on mode identifications
is the mode density. As mentioned in \S 2.3.1, models predict roughly
one overtone ($n$) per degree ($\ell$) per 1000~$\mu$Hz and 
though we have detected 28 frequencies, they are
spread across nearly 6000~$\mu$Hz. Since we do not detect any multiplets
that can be unambiguously associated with rotational
splitting, it is likely the pulsations are degenerate in $m$.
%$m=0$, though it is certainly possible that some of the splittings near
%41.1~$\mu$Hz are actual multiplets associated with $m\neq 0$ modes.
The average mode density
is 4.8 frequencies per 1000~$\mu$Hz, which is too high to accommodate
only $\ell\leq 2$, with degenerate $m$ modes and it gets worse as 
the frequencies are not quite
distributed equally, but rather fall into loose groups, enhancing  the
density locally. The regions between 5200 and 6200
and 6600 to 7600~$\mu$Hz each contains 9
frequencies, and the region from 8800 to 9800~$\mu$Hz has 5 frequencies.
%all of which are
%too many to be accounted for using only low-degree modes. 
If all $2\ell +1$ multiplets were filled (9 frequencies per 1000~$\mu$Hz),
the frequency density would not require any $\ell \geq 3$ modes. However,
as we do not detect appropriate multiplets within these frequency regions,
the most likely result is that PG~0048 has too high a frequency
density to exclude $\ell\geq 3$ modes using current models.

\begin{figure}
\figurenum{11}
\includegraphics[width=\textwidth]{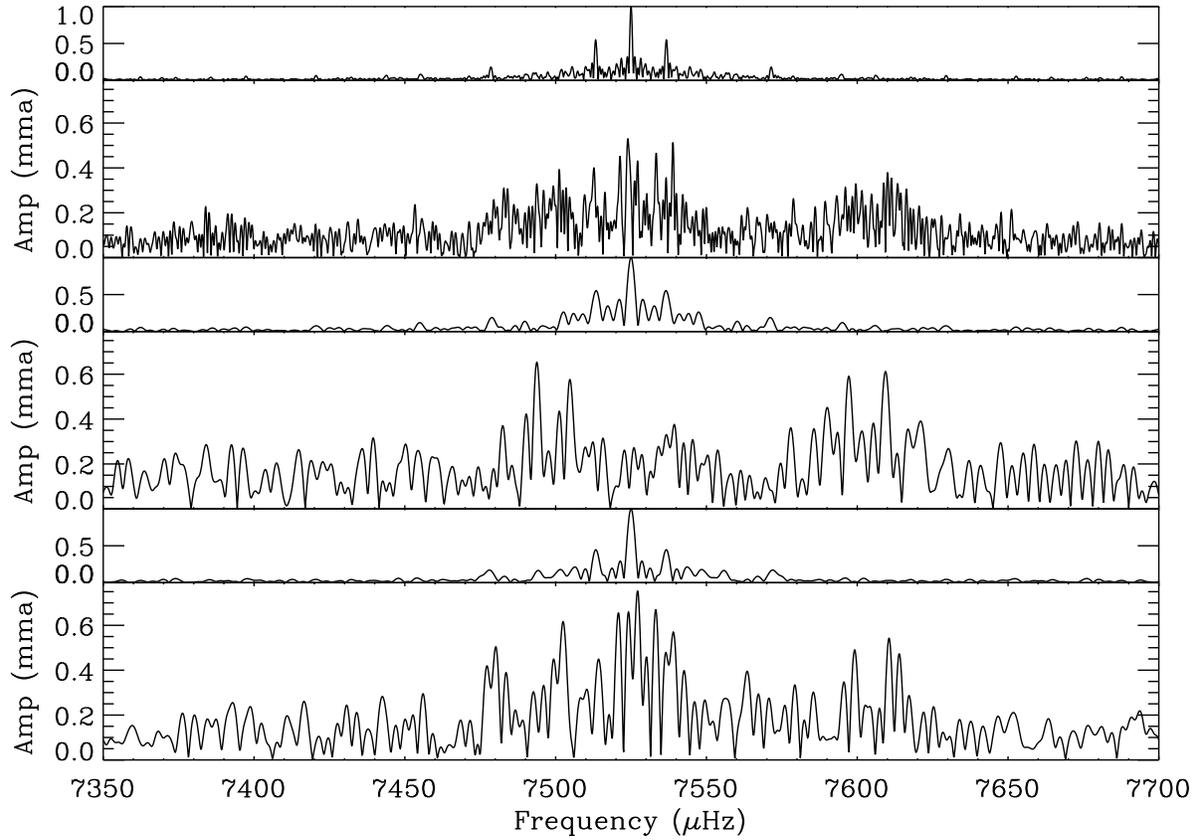}
\caption{A 350\,$\mu$Hz region of PG~0048's FT covering three modes at
  $\sim$7497.4\,$\mu$Hz, $\sim$7524.1\,$\mu$Hz and
  $\sim$7605.9\,$\mu$Hz. The panels correspond to those in 
Fig.~\ref{fig06}. The spectral windows at the top of each panel cover the
  same frequency range as the observations, to give the reader an
  appreciation of how dense these regions are.}
\label{fig11}
\end{figure}

\placefigure{fig11}

\subsubsection{Amplitude and phase variability}

Nearly all sdBV stars show some amount of amplitude variability.
However, no other sdBV star has shown variability like that detected in PG~0048.
An example of this is shown in Fig.~\ref{fig11}, where
the same 350~$\mu$Hz region is shown for the three data sets in
Fig.~\ref{fig06}. The bottom two panels are subsets of the top panel, 
which includes all of the data, and indicate how strikingly the
pulsation spectrum changes with time.

\begin{figure}
 \figurenum{12}
 \includegraphics[angle=-90,width=\textwidth]{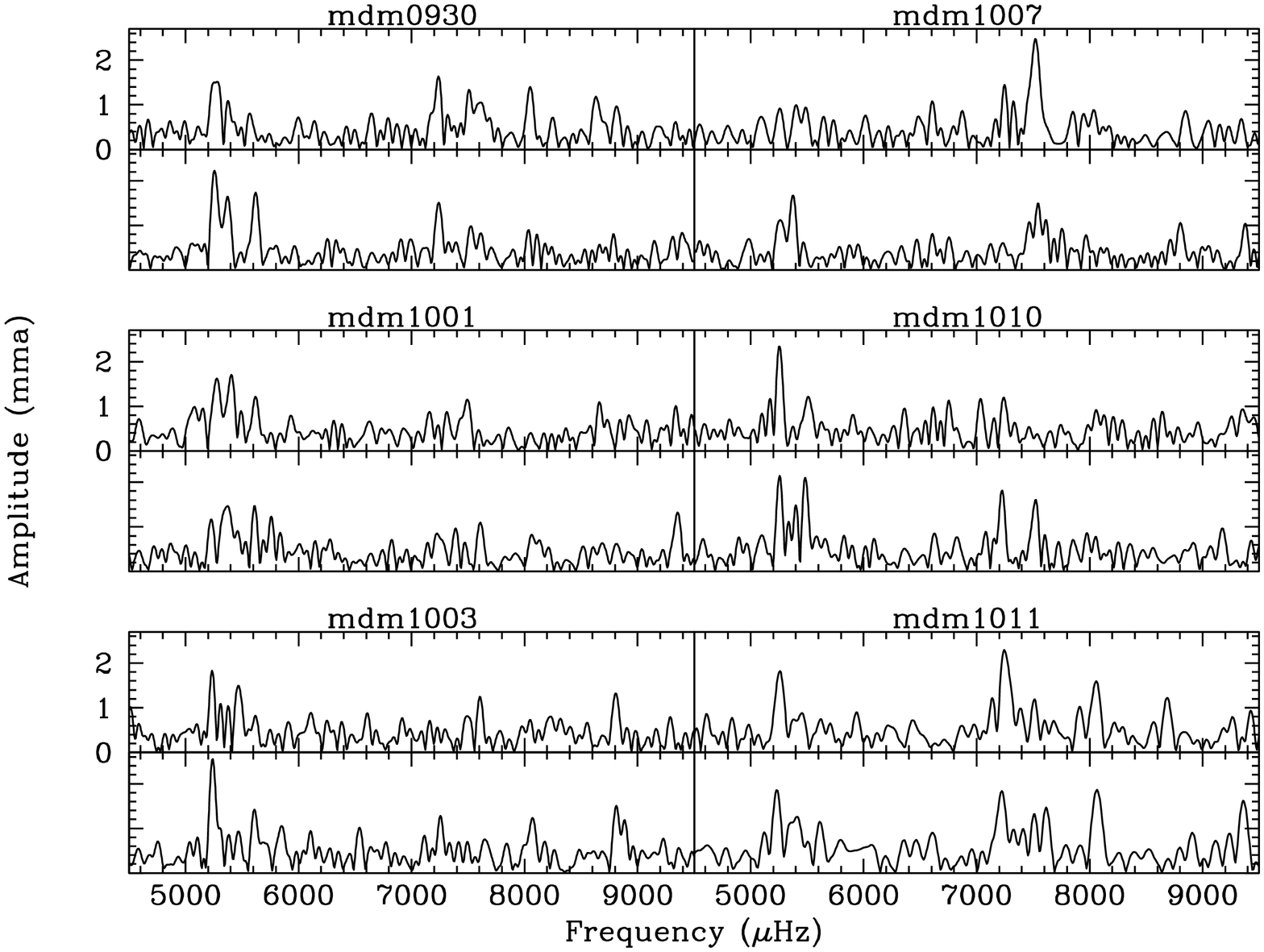}
 \caption{
 Temporal spectra for halves of individual PG~0048 observing runs.}
 \label{fig12}
 \end{figure}

\begin{figure}
\figurenum{13}
\includegraphics[angle=-90,width=\textwidth]{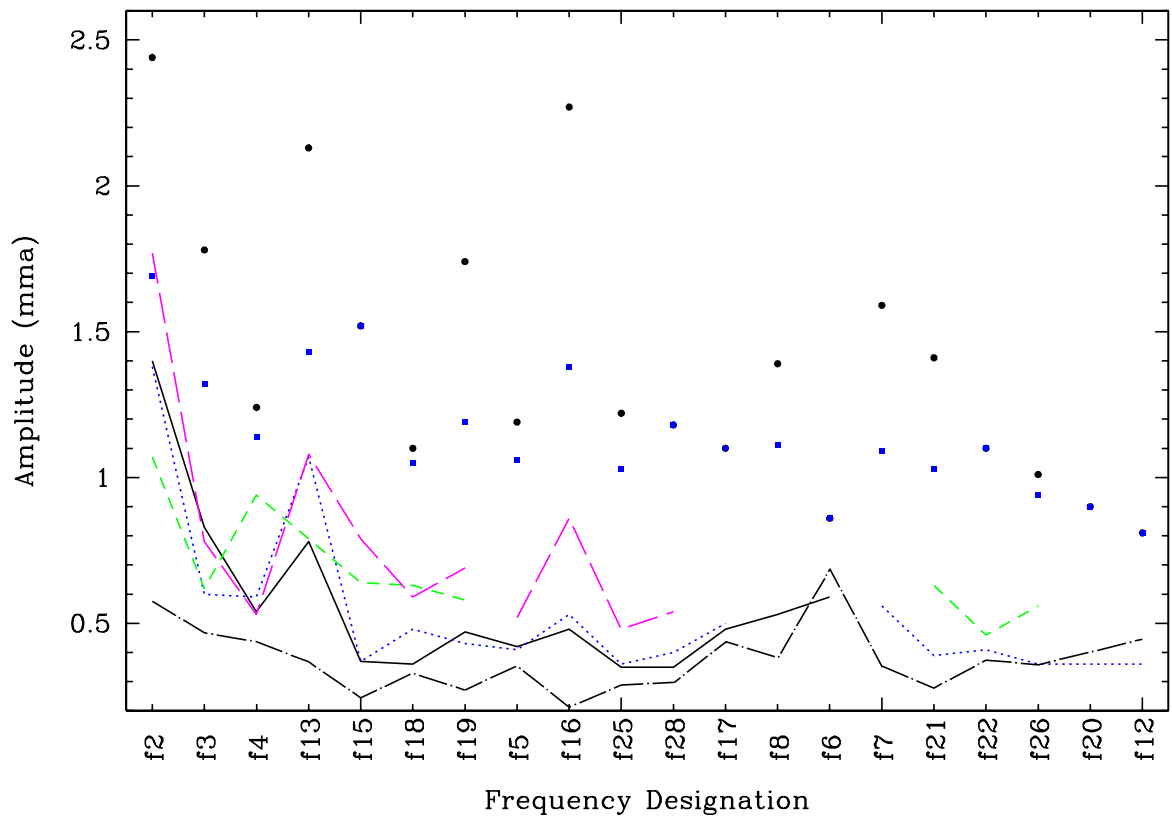}
\caption{Amplitudes of frequencies detected in both individual runs and
grouped data during 2005 for PG~0048.
Dots (black) indicate the maximum amplitude, squares (blue) the average
amplitude from individual runs. The lines indicate amplitudes from the following
group datasets (from Table~\ref{tab06}): dotted (blue) is G1, short-dashed (green) is G2, long-dashed
(magenta) is G3, solid (black) is G4, and the dot-dashed (black) line is
the $G4/A_{\rm max}$ ratio (or $G1/A_{\rm max}$ if no G4 value).}
\label{fig13}
\end{figure}

% {\b CHOOSE BETWEEN THE FOLLOWING TWO DISCUSSIONS: DISCUSSION 1:}
 We can apply some constraints to the timescale of
 amplitude variability. Since pulsation frequencies can change amplitudes
 (even to the point of being undetectable) between individual runs,
 and in particular between runs from different observatories but for the
 same date, the
 next step was to divide up our longer data runs into halves. The FTs
 for six such runs are shown in Fig.~\ref{fig12} with each half containing
 about four hours of data.
% 
% {\b DISCUSSION 2: Don's preference} 
%We can apply some constraints to the timescale of
%amplitude variability. Since pulsation frequencies can change amplitudes
%(even to the point of being undetectable) between individual runs,
%and in particular between runs from different observatories but for the
%same date, the
%next step was to divide up our longer data runs into smaller temporal
%segments. We created three subsets of data from our best individual runs in
%the following manner: The first and last half of the data and the central
%portion of equal length to each half. In this manner, we have a temporal
%slice across the run in three segments, each with the same temporal resolution,
%though the central slice has the benefit of the lowest point-to-point
%noise (because of the lower airmass). Adjacent temporal
%slices share half their data points with each slice containing
%about four hours of data. The FTs
%for four such runs are shown in Fig.~\ref{fig12} with the $4\sigma$ detection
%limit indicated by the dashed lines.
% 
% {\b BACK TO THE NORMAL TEXT} 
Particularly for runs mdm1007  (which can
also be compared to saao1007) and mdm1010, the amplitudes change by factors of
two over times as short as four hours. Low amplitude frequencies
can easily become undetectable 
within that time.  Figure~\ref{fig13} compares the maximum
and average amplitudes detected in individual runs (the black circles
and blue squares, respectively) to detections in groups of data (the
lines) from Table~\ref{tab06}. 
If the amplitudes were simply wandering around
between values detected in individual runs, then the amplitude of the
combined data would be an average of these values (the blue squares).
However, since the combined
amplitudes are significantly lower then the average amplitudes from individual
runs, something else must be occurring to reduce the amplitudes.

\placefigure{fig12}

\placefigure{fig13}

\begin{figure}
\figurenum{14}
\includegraphics[angle=-90,width=\textwidth]{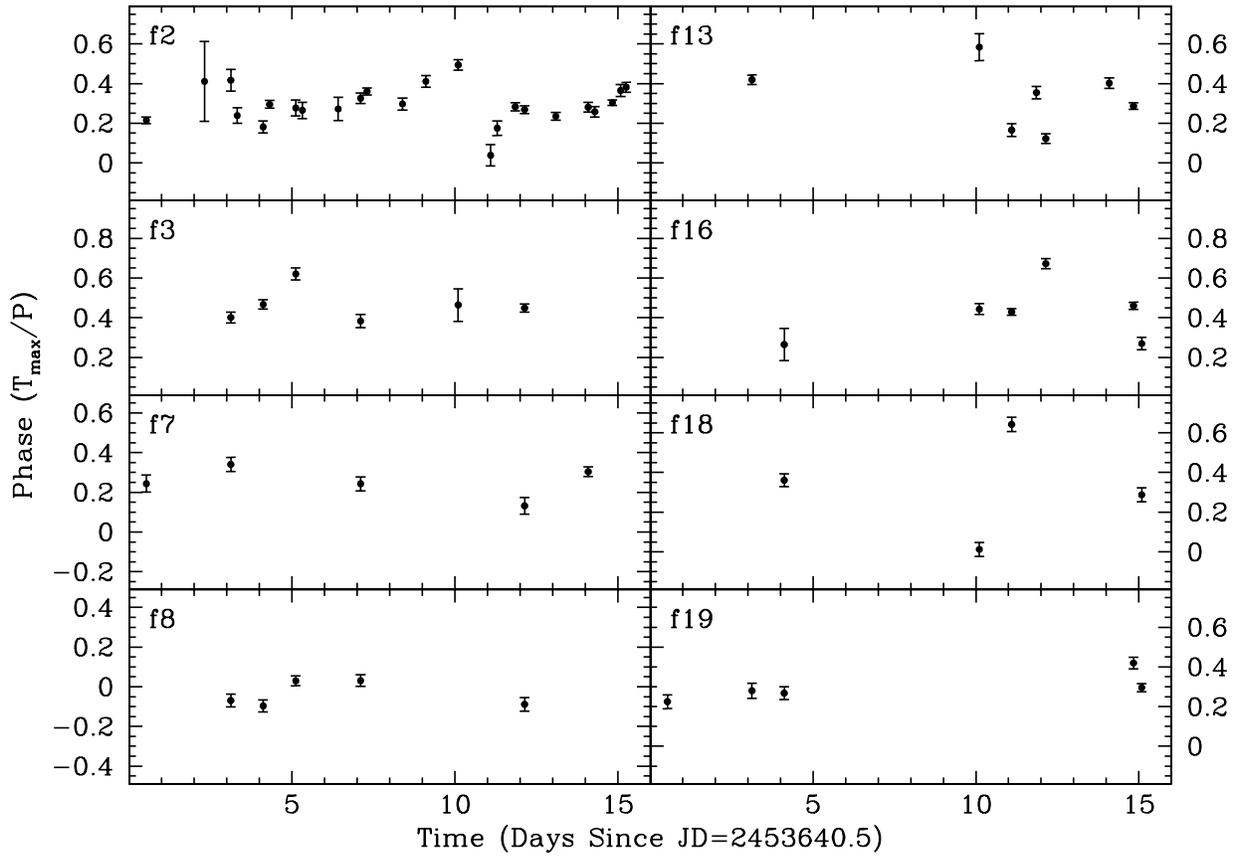}
\caption{Phases of pulsation frequencies. Frequency designations are provided
in each panel with phases determined only for individual runs in which the
frequency was detected.}
\label{fig14}
\end{figure}

Since changes in phase can impact pulsation amplitudes, we investigated
that next. Figure~\ref{fig14} shows phases for eight of the most-often
detected frequencies. $f2$ is the only frequency detected in every run, and
we include the half-night analysis for it. For the other seven frequencies, we
only determined phases for those runs in which they were detected above the 
$4\sigma$ limit. While Fig.\ref{fig14} indicates that most phases
do not appear constant with time, most are within 20-30\% of a central
value. 

To isolate and test the impact phase variation creates on data like ours,
we analyzed simulated data with the following properties: The data are
represented by a noise-free, single-frequency sine wave sampled
corresponding to our 12 best individual runs with frequency $f2$. The
amplitudes of each individual run are fixed at the measured values for 
$f2$ from Table~\ref{tab06} and we assume that \emph{no} phase changes
occur during an individual run. While the properties of $f2$ may not be
the most representative of the variations detected, it is the only 
frequency detected every time, and so by using it, we sample the full range
of amplitudes. If we used a different frequency, we could not know the
actual pulsation amplitude during those runs without detections, and so would
only be sampling the higher amplitude data points. 
We created simulations with the following
phase properties: No change in phase, a fixed change of $\pm 10$\% from the 
previous phase, a fixed change of $\pm 20$\% from the previous phase,
phases that are randomly set at the beginning of each run, and with the
actual phase values for $f2$. The results of the simulations are given in
Table~\ref{tab09} and the ratios are used in
Fig.~\ref{fig15} where they are compared with our 
observations. The last line of Table~\ref{tab09} presents results with
unique answers.
As expected, there is a correlation between the amplitude
and the amount of phase variation in that increasing changes in phase
between individual runs decreases the measured amplitude of the data set
as a whole. More useful are the ratios of the average amplitude to the
average and maximum of the individual amplitudes ($\langle A\rangle/\langle 
A\rangle_{ind}$ and $\langle A\rangle/A_{max}$). These ratios can be
compared with ratios from all frequencies, as has been done in 
Fig.~\ref{fig15}. The shaded regions are the $1\sigma$ ratios produced from
the phase simulations and the circles represent ratios with $G4$ data and
the squares are for ratios using $G1$ data. The frequency ordering is
the same as for Fig.~\ref{fig12}, and like Fig.~\ref{fig12},  the results 
indicate that for all but $f2$ and possibly $f6$, 
the amplitudes detected in groups of
data are too low compared to individual amplitudes. Additionally, except
for $f2$, the phases of Fig.~\ref{fig14} (and their standard deviations
given in Table~\ref{tab10}) are in discord with the amplitude ratios in
that the amplitudes are too low. There does not seem to be sufficient
phase variation to produce the low amplitudes of the group data. As such,
it seems that more extreme circumstances are required. However, that
leads into a more speculative area which we save for \S 3.3.4.

We conclude our observational portion of this paper with a summary provided
in Table~\ref{tab10}. In this table we have included all measurables (not
in previous tables) discussed in this section and some that will be useful
for the next section. Columns 2, 3, and 4 consider the number of expected
detections based on the average significance of the actual detections;
Columns 5 through 8 detail amplitudes detected for individual 2005
observing runs; Columns 9 through 12 provide ratios of individual to
group amplitudes; and Column 13 lists the $1\sigma$ deviations of pulsation
phases.

\placefigure{fig14}

\subsubsection{A possible cause of the amplitude/phase variability}

Can we determine the cause of the erratic behaviour of
PG~0048's frequencies and amplitudes? We suggest that,
with the possible exception of $f2$, the oscillations may be
stochastically excited.
% the oscillations are randomly excited, with the possible exception of
While this is counter to current theory, supporting
observations that we have in hand are 
1) amplitudes that vary
significantly between \emph{every} individual run, and in less than
$\sim$4\,hrs; 2) the combined amplitudes are significantly lower than the
average value indicating that phases are not coherent on these
timescales; 3) the peaks in the FTs appear similar to those of 
known stochastic pulsators (compare Fig.~\ref{fig11} to Fig.~1 of
Bedding et al. (2005) or Fig.~2 of Stello et al. (2006) for stochastic
oscillators to those in, for example, Reed et al (2004) for 
``normal'' sdB stars); and 4) the number of actual frequency detections
compared to the expected number based on significance 
(the ratio of the two lines in the left panel of Fig.~\ref{fig09}). 
We note that this is not conclusive evidence, but is
suggestive and so we will pursue a stochastic nature for PG~0048's
pulsations in the remainder of this section.

Recently, Stello et
al. (2006) found that short mode lifetimes in red giants can severely limit
the possibility of measuring reliable frequencies. The difficulty arises
because the frequencies can disappear entirely and when they are
re-excited (even if this occurs prior to complete damping), they do not
maintain the same phase. The parallel with our analysis of PG~0048 are
clear and this kind of variability has been seen before in sdBVs.
In a study of KPD\,2109+4401, Zhou et al. (2006) found
substantial variation in the amplitudes of two modes during their
32 night campaign. A brief analysis found that at least one of these
modes, and possibly both of them, satisfied the criterion outlined by
Christensen-Dalsgaard et al. (2001; hereafter JCD01) for
stochastically excited pulsations, rather than overstable
driving. The criterion compares the ratio of amplitude scatter to the
mean amplitude; for stochastic pulsations, this ratio should be $\approx
0.52$. Stochastic processes in pulsating sdB stars
have also been discussed by Pereira \& Lopes (2005) in the context of
the complex sdB pulsator PG\,1605+072, which is known to have variable
amplitudes (O'Toole et al. 2002; Reed et al. 2007, in press). 
Using the JCD01 criterion, Pereira \&
Lopes deduced that none of the modes of that star were consistent with
stochastic excitation. However, O'Toole et al. (2002) noted amplitude 
changes between years, while Pereira \& Lopes (2005) only studied 7 
nights of data, and as such their analysis was likely affected
by the short length of their time series. 

A limitation to the JCD01 test is that the damping
times of the oscillations should be longer than the timescale used 
for determining the amplitudes. Our analysis of $\sim 4$ hour segments of
PG~0048 data indicate that amplitude variations are on very
short timescales that are shorter than the observing time
for individual runs (see Figs.~\ref{fig07} and \ref{fig12}). We
provide the JCD01 parameter values $\sigma_A/\langle A\rangle$ in Column
7 of Table~\ref{tab10}, but
%This
%will smooth out the variations, which will reduce the amplitude
%scatter. We cannot divide the observations into smaller temporal
%blocks since we would quickly lose the ability to resolve the
%pulsations, as well as a fair amount of frequency precision. 
we suggest that the JCD01 test is not appropriate for PG~0048.
Aside from the JCD01 test, we can attempt to reproduce some of the
observational properties using simple simulations
with damped and randomly re-excited frequencies.
The complexity of the actual data is such that we cannot hope to
reproduce it directly, but instead will strive to fit the 
observations listed at the beginning of this section.
% al constraints:  1) The ratio of
%the maximum amplitude observed in a single run and the 
%number of detections from individual runs 
%(Fig.~\ref{fig14}); 2) amplitudes that decrease by half within
%four hours; and 3) the ratio of observed amplitudes from the
%entire data set to the maximum observed. 
Our simulations follow
the simple prescription (equations 2 and 3) of Chaplin et al.
(1997) summarized as follows: The pulsations themselves are
described by sine waves of the form 
$A(t)=A\cos((2\pi\cdot f)(t-\phi ))$, with the amplitude modified in
two ways; it is damped exponentially as $A=A_o\exp (-t_e/t_d)$ where
$A_o$ is the maximum amplitude,
$t_e$ is the time since the last excitation and $t_d$ is the 
damping timescale. The pulsation is re-excited by
setting $t_e=0$ when time
$t$ exceeds an excitation timescale ($t_{exc}$).
The time before the first
re-excitation is randomly set to some fraction of the excitation
timescale and every time the pulsation is re-excited,
the phase is randomly set and the excitation timescale and
pulsation amplitude can vary randomly by up to 20\% or their 
original values.
The free parameters of the simulation are the input amplitude, which is
the maximum amplitude attainable  and the excitation
and damping timescales ($A_o\,,t_{exc}\,$ and $t_d$, respectively). 
The simulation includes frequencies,
amplitudes, and phases for up to 100 pulsations with
an unlimited number of data runs (input as run start time, the number of
data points, and cycle time). 
We will concentrate on matching the MDM runs in Table~\ref{tab06}.
This is the $G1$ data set with an average run length of 9.35 hours
($\sim 0.4d$), containing $\sim 2800$ data points each, an average
detection limit of 0.88~mma, and average ratios $A_{\rm G1}/\langle A\rangle
=0.44\pm0.14$ and  $A_{\rm G1}/A_{\rm max}=0.37\pm0.09$.

To match the observational
constraint that the pulsation amplitude can reduce by half in a four
hour span, the damping timescale is necessarily less than 5.8 hours. 
With this constraint, we produced 
a grid of simulations with $1\leq t_d \leq7$ hours in 1 hour increments and 
$1\leq t_{exc} \leq33$ hours in 2 hour increments. Qualitatively,
if no re-excitations occur during an individual run, the FT is single-peaked
whereas multiple re-excitations create a variety of complex, multi-peaked
FTs, depending on how similar the randomized phases were (the 
less alike the phases for each re-excitation, the lower the overall
amplitude and more and similar-amplitude peaks appear in the FT).
For small
$t_{exc}$, the FT becomes increasingly complex with most simulations
resulting in many low-amplitude peaks distributed across a couple hundred
$\mu$Hz. However, such complex patterns are not consistent with observations
and so we discount small values for $t_{exc}$. Amplitudes in the FT are
reduced with large values of $t_{exc}$ and small values of $t_d$ while
the scatter increases with increasing values for both. An increase in 
amplitude scatter is necessary to produce the low rate of detections.

% Roughly, the results
% of single run data simulations are that any 
% frequency with an intrinsic maximum amplitude
% twice the detection limit (1.76~mma for the MDM data) should be 
% detected in \emph{all} MDM data runs if 2~hr~$\leq t_d\approx t_{exc}$ or
% 4~hr~$\leq t_d\leq$~9.35~hr (the run length). From our observations,
% there are four frequencies that minimally fit this amplitude criteria
% (i.e. they are detected to have amplitudes $\geq 1.76$~mma during at least
% one data run) yet only one of them is actually detected in all runs.
% As such, we need to search for longer re-excitation timescales to lower
% the overall average detection amplitude.

\begin{figure}
\figurenum{16}  
\includegraphics[angle=-90,width=\textwidth]{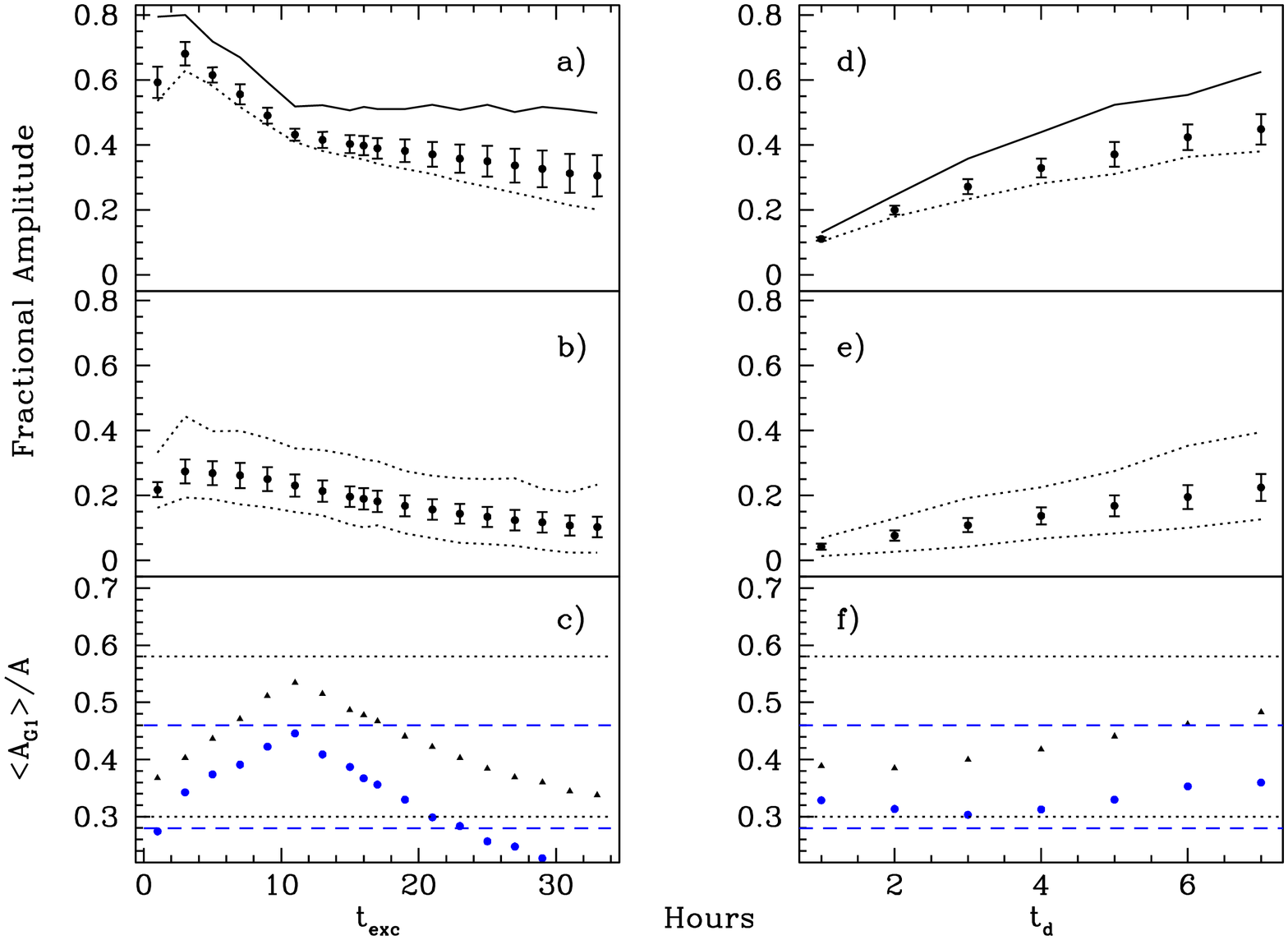}
\caption{Simulations of randomly excited and damped pulsations.
Panels a, b, and c show the effect of changing the excitation timescale
for a fixed damping timescale of $t_d=5$~hours. Panel a and b show the results of
simulating a single 9.35 hour run and the nine MDM runs in Table~\ref{tab02},
respectively. The points indicate the average amplitude with $1\sigma$
error bars, the solid line is the maximum amplitude from an individual
simulation and the dotted line indicated the minimum amplitude
detected from an individual simulation. Panel c shows the ratios 
$\langle A_{G1}\rangle/A_{max}$ (blue circles) and 
$\langle A_{G1}\rangle/\langle A_{ind}\rangle$ (black triangles).
The dashed blue (dotted black) lines indicate the observed range in PG~0048
for $\langle A_{G1}\rangle/A_{max}$ 
($\langle A_{G1}\rangle/\langle A_{ind}\rangle$). Panels d,
e, and f correspond to panels a, b, and c except that the excitation
timescale is fixed at $t_{exc}=19$~hours and the damping timescale is varied.}
\label{fig16}
\end{figure}

\placefigure{fig16}

\placefigure{fig17}

% The simulations of the combined MDM data set are shown in panels b and e of 
% Figure~\ref{fig15} and indicate, as expected, lower amplitudes
% than in the individual runs. Again, this is caused by the randomness of
% the phases, which results in partial cancellation of amplitudes. The
% bottom panels of Figure~\ref{fig15} show the ratio of the
% average amplitude seen in the combined data set to the highest
% amplitude detected in an individual run, $\langle A\rangle /A_{max}$.
% These should be compared to the $A_{MDM}/A_{max}$ ratio provided above
% and the dashed blue lines of the figure indicate the $1\sigma$ limits
% of that value. 

As we now have all the pieces in place, we can ask how
well the simulations reproduce the observational constraints
we set at the beginning of this section. A selection of the
results are shown in Figs.~\ref{fig16} and \ref{fig17}.
Panels a, b, and c of Fig.~\ref{fig16} have $t_d$ fixed at 5 hours and vary
$t_{exc}$ whereas panels d, e, and f fix $t_{exc}$ at 19 hours and vary $t_d$.
Panels a and d are the results for individual runs and panels b and e are for
the combined nine-run data set. The points represent the average ($\langle A\rangle$)
with $1\sigma$ deviations while the lines indicate maximum ($A_{max}$) and 
minimum amplitudes. Panels c and f show the ratios $\langle A_{G1}\rangle
/A_{max}$ and $\langle A_{G1}\rangle/\langle A_{ind}\rangle$; the average
amplitude from the combined data divided by the maximum or average amplitude
from the individual runs. Figure~\ref{fig17} shows the expected rate of detections
calculated in the following manner: The detection limit was calculated using 
the observed ratios $\langle A_{max}\rangle/0.88=1.56$ and 
$\langle A\rangle/0.88=1.32$ for the MDM data with an average detection limit
of 0.88~mma and solving for the new detection limit.
The dotted line is the observed detection rate of 26\% and it is 
interesting that none of simulations are that low if using the average
detected amplitude. While the overall amplitudes can become quite small, the
average follows that, which is used to calculate our detection limit in the 
top panel. However, most values of $t_d$ matched the observed rate near
$t_{exc}=19$-$21$~hours using $A_{max}$.
% We have already ruled out
% $t_d> 6$~hrs and small $t_{exc}$ and the simulations agree with that
%interpretation: Though $\langle A\rangle /A_{max}$ for 
The simulations that
best fit the observational constraints are those which have
$4\leq t_d\leq 6$~hrs and $13\leq t_{exc}\leq 21$~hrs. The lower values
of $t_{exc}$ better fit the $\langle A_{G1}\rangle/A$ ratios while the
larger values are a better match for the detection rate. These
relatively simple simulations are able to fit all of the observed constraints,
thus explaining the amplitude variations, their lower detection limits in
groups of data, the relatively low detection rate and the appearance of the
peaks in the FT. What they cannot explain however, is the relative lack
of phase variability in some frequencies (though relatively few were measurable)
and \emph{why} stochastic processes should occur in the first place.

Stochastic oscillations are usually presumed to be driven by random
excitations caused by convection (see Christensen-Dalsgaard 2004 for
a review concerning the Sun). The
He~II/He~III convection zone in sdB stars was investigated by 
Charpinet et al. (1996), who determined that it could not drive
pulsations. However, it appears that they did not investigate this zone for
convective motions but rather as a driver for the
$\kappa$ mechanism. So some ambiguity remains here.
%While we acknowledge that theoretical models do not
%show any substantial convection zones near the surface of sdB stars,
%observational evidence of stochastically driven modes would
%be tremendously exciting. 
There is also convection or semi-convection in
the cores of sdB stars, and it is possible that the eigenfunctions
could be sampling this region. Whatever the case, the
extreme amplitude and phase variability of PG~0048 poses a significant
challenge to the iron driving mechanism found by Charpinet et
al. (2001) to excite pulsations in sdB stars. Though it is beyond the
scope of this paper
to ascertain the \emph{cause} of the random amplitude variations,
we find that the observed properties are consistent with 
our simplified randomly excited simulations and that the amplitude
spectrum resembles those of pulsators that are stochastically driven.

\section{Conclusions and Future Work}

We have carried out multisite campaigns for two sdB pulsators,
PG~1618+563B and PG~0048+091 and %revealed the great differences
in both cases, our observations were
superior to the published discovery data, yet questions concerning
these two stars still remain.
Our MDM observations of PG~1618B (obtained
under good conditions) show
characteristics typical of about half the objects in the sdBV
(V361\,Hya) class: A small number of stable (in amplitude) frequencies
with a closely spaced pair. In contrast, the data obtained at McDonald
observatory -- under non-photometric conditions -- show PG~1618B to be a
complex pulsator with four ``regions'' of power showing amplitude and phase
variability. An ensemble analysis of any combinations of data other than
the MDM set are hindered by poor least-squares fitting and amplitudes reduced below
detectability. Such poor fitting can be caused by unresolved frequencies
with intrinsic amplitude variability (O'Toole et al. 2002) or
randomly excited pulsations (Christensen-Dalsgaard 2004).
 So despite having expended
considerable effort to obtain not only multisite, but extended
time-base observations, we were only reasonably successful at
detecting pulsations from our 2~m telescope data; and these show the
star to be two-faced. With this dichotomy of observational results,
PG~1618B %(somewhat disappointingly) SOUNDS VERY NEGATIVE!!
remains an interesting target for more follow-up observations; particularly
to examine its long-term frequency stability.

PG~0048 is much more complex than PG~1618B, yet it too has
shown somewhat stable pulsation amplitudes at one epoch (the 
discovery and 2004
data) and wildly variable amplitudes at another (2005). Though an 
extremely rich pulsator with at least 28 independent frequencies,
many modes are only excited to amplitudes above the noise
occasionally, often for very short lengths of time.
These behaviors are consistent with stochastic
pulsations and we have performed several tests along these lines. We
simulated damped and re-excited pulsations and found that the observations
were best matched with damping timescales between 4 and 6 hours and excitation
timescales between 13 and 19 hours. We detected common frequency splittings
of 972 and 41~$\mu$Hz which may be related to multiplet structure, but
could reproduce these using Monte Carlo splittings of random spacings. So
while they may be intrinsic to the pulsation of PG~0048, we cannot be sure.
We can be sure that PG~0048's rich pulsation spectrum is too
dense to be accounted for using only $\ell\leq 2$ modes regardless of how
many $m$ components are present.

The observations presented in this paper provide some very interesting
and confusing results. Pulsations that appear stable during some times and
variable at others; attributes that have also been observed
in other sdBV stars as well (KPD~2109+2752, PG~1605+072, and 
HS~1824+5745, just to name a few). The pulsations in PG~0048 present
observables that seem best
described by randomly excited oscillations which would be in contrast
to the proposed driving mechanism (Charpinet et al. 2001). If validated, 
it would represent
a new direction in sdB pulsations (and modeling too!). However, a
longer time series may be the only way to clarify the nature of the
oscillations in this star. Ideally this would take place on at least
2\,m-class telescopes and cover several weeks. 

\acknowledgments
We would like to thank the time allocation committees for 
generous time allocations, without which
this work would not have been possible; Dave
Mills for his time and help with the Linux camera drivers; Chris
Koen for contributing the discovery data and helpful discussions; and
the anonymous Referee for a helpful re-organization of the paper.
This material is based in part upon work supported by the 
National Science Foundation under Grant Numbers AST007480.
Any opinions, findings, and conclusions or recommendations
expressed in this material are those of the author(s) and do not necessarily
 reflect the views of the National Science Foundation.
JRE, GAG, and SLH were supported by the Missouri Space Grant Consortium.

\clearpage

%Table 1
\begin{table}
\tablenum{1}
\centering
\caption{Observation record for PG~1618B. The first two runs were obtained in
2003, while the rest were obtained in 2004. \label{tab01}}
\begin{tabular}{lccccc} \hline 
Run & Date & Start & Length & Int. & Observatory \\  
& UT & hr:min:sec &  (Hrs) & (s) &  \\ \hline
suh16mar & 17 Mar & 00:07:58 & 3.3 & 10 & Suhora 0.6~m \\
lul031705 & 17 Mar & 16:46:33 & 5.6 & 10 & Lulin 1.0~m \\
McD031805 & 18 Mar & 04:34:00 & 1.1 &  5 & McDonald 2.1~m  \\
lul031805 & 18 Mar & 15:14:24 & 6.1 & 10 & Lulin 1.0~m \\
McD031905 & 19 Mar & 08:40:40 & 3.1 &  5 & McDonald 2.1~m  \\
lul031905 & 19 Mar & 19:58:41 & 1.3 & 15 & Lulin 1.0~m \\
baker032005 & 20 Mar & 04:29:30 & 5.6 & 25 & Baker 0.4~m \\
McD032005 & 20 Mar & 06:35:00 & 5.8 &  5 & McDonald 2.1~m  \\
lul032005 & 20 Mar & 15:17:19 & 6.0 & 10 & Lulin 1.0~m \\
suh20mar & 20 Mar & 18:59:00 & 8.3 & 10 & Suhora 0.6~m \\
lul032105 & 21 Mar & 16:44:59 & 0.7 & 15 & Lulin 1.0~m \\
suh21mar & 21 Mar & 18:20:20 & 8.6 & 20 & Suhora 0.6~m \\
McD032205 & 22 Mar & 04:33:00 & 7.8 &  5 & McDonald 2.1~m  \\
suh22mar & 22 Mar & 18:40:40 & 2.2 & 20 & Suhora 0.6~m \\
McD032305 & 23 Mar & 04:12:10 & 8.0 &  5 & McDonald 2.1~m  \\
mdr299 & 29 Mar & 04:11:15 & 4.7 & 15 & Baker 0.4~m \\
mdr301 & 31 Mar & 04:28:10 & 7.0 & 15 & Baker 0.4~m \\
mdr302 & 02 Apr & 02:53:10 & 8.1 & 10 & Baker 0.4~m \\
bak040305 & 03 Apr & 03:20:46 & 7.9 & 15 & Baker 0.4~m \\
bak040405 & 04 Apr & 04:23:10 & 5.6 & 15 & Baker 0.4~m \\
bak040505 & 05 Apr & 03:17:10 & 6.6 & 10 & Baker 0.4~m \\
bak041405 & 14 Apr & 03:45:00 & 7.1 & 10 & Baker 0.4~m \\
bak041505 & 15 Apr & 02:26:30 & 8.4 & 10 & Baker 0.4~m \\
bak041605 & 16 Apr & 02:50:30 & 7.9 & 10 & Baker 0.4~m \\
bak041705 & 17 Apr & 03:18:50 & 5.3 & 15 & Baker 0.4~m \\
bak041805 & 18 Apr & 02:56:00 & 7.7 & 15 & Baker 0.4~m \\
mdm042605 & 26 Apr & 04:25:30 & 7.3 &  5 & MDM 2.4~m  \\
mdm042705 & 27 Apr & 04:16:50 & 7.6 &  5 & MDM 2.4~m  \\
mdm042805 & 28 Apr & 04:14:00 & 7.8 &  3 & MDM 2.4~m  \\
mdm042905 & 29 Apr & 04:18:00 & 1.6 &  5 & MDM 2.4~m  \\
mdm043005 & 30 Apr & 04:04:30 & 7.8 &  3 & MDM 2.4~m  \\
mdm050105 & 01 May & 04:05:30 & 7.8 &  5 & MDM 2.4~m  \\
mdm050205 & 02 May & 03:36:40 & 5.3 &  5 & MDM 2.4~m  \\ \hline
\end{tabular}
\end{table}

\clearpage

\begin{table}
\tablenum{2}
\caption{Subsets of data for PG~1618B. For column 2, observatories are 1) 
McDonald Observatory; 2) Suhora Observatory; 3) Lulin Observatory; 4) Baker
Observatory; and 5) MDM Observatory.}
\begin{tabular}{lcccc}
\hline
Set & Observatory(ies) & 
Inclusive Dates & Resolution & $4\sigma$ detection limit \\ 
   &  &  & $\mu$Hz & mma \\ \hline
McD & 1 & 18 - 23 March & 2.2 & 1.64 \\
MDM & 5 & 26 Apr - 02 May & 1.9 & 0.55 \\
McD+MDM & 1, 5 & 18 - 02 May & 0.3 & 0.59 \\
Week 1 & 1, 2, 3 & 17 - 23 March & 1.5 & 1.53 \\
All & 1, 2, 3, 4, 5 & 17 - 02 May & 0.2 & 0.77 \\ \hline
\end{tabular}
\label{tab02}
\end{table}

\clearpage

\begin{table}
\tablenum{3}
\caption{Our least-squares fit solution for the pulsation periods,
frequencies, and amplitudes detected in PG~1618B. Formal least-squares
errors are given in parentheses. Frequencies marked with a dagger 
($^{\dagger}$) were only detected during individual runs (one each) while the 
remaining frequencies are from the MDM data set.}
\begin{tabular}{lccc}
\hline
Des. & Period & Frequency  & Amplitude  \\ 
& (s) & ($\mu$Hz) & (mma) \\ \hline
$f1^{\dagger}$ & 108.7092 (0.0518)  &  9198.85 (4.38) & 1.79 (0.39) \\ 
$f2^{\dagger}$ & 122.2574 (0.0597)  &  8179.46 (4.00) & 1.04 (0.21) \\ 
$f3$ & 128.9549 (0.0008)  &  7754.64 (0.05) & 1.71 (0.09)\\
$f4$ & 139.0571 (0.0008)   & 7191.28 (0.04) & 2.04 (0.09)\\
$f5$ & 143.9290 (0.0011) &   6947.87 (0.05) & 2.22 (0.10)\\
$f6$ & 143.9759 (0.0014)   & 6945.60 (0.07) & 1.64 (0.10) \\ \hline
\end{tabular}
\label{tab03}
\end{table}

\clearpage

%Table 4
\begin{table}
\tablenum{4}
\centering
\caption{Observations of PG~0048 \label{tab04}}
\begin{tabular}{lccccc} \hline
Run & Date & Start & Length & Int. & Observatory \\
 & UT & hr:min:sec &(Hrs) & (s) &  \\ \hline
\multicolumn{6}{c}{2004} \\
mdr285 & 10 Oct & 04:21:00 & 6.2 & 15 & MDM 1.3~m\\
mdr290 & 12 Oct & 03:39:00 & 7.3 & 15 & MDM 1.3~m\\
mdr295 & 14 Oct & 03:34:00 & 7.4 & 15 & MDM 1.3~m\\ \hline
\multicolumn{6}{c}{2005} \\
boao & 26 Sept. & 10:30:20 & 5.0 & 10 & BOAO 1.9~m  \\
mdm092805 & 28 Sep & 07:32:30 & 4.5 & 15 & MDM 1.3~m  \\
mdm092905 & 29 Sep & 02:53:00 & 9.5 & 15 & MDM 1.3~m   \\
mdm093005 & 30 Sep & 02:44:00 & 9.5 & 15 & MDM 1.3~m   \\
turkSep3sdb & 30 Sep & 17:37:54 & 9.3 & 10 & Tubitak 1.5~m \\
mdm100105 & 01 Oct & 02:46:00 & 9.4 & 10 & MDM 1.3~m  \\
turk1Octsdb & 01 Oct & 21:26:58 & 4.1 & 10 & Tubitak 1.5~m \\
mdm100205 & 02 Oct & 09:57:30 & 2.3 & 15 & MDM 1.3~m  \\
turk2Octsdb & 02 Oct & 20:51:36 & 4.8 & 10 & Tubitak 1.5~m \\
mdm100305 & 03 Oct & 02:32:00 & 9.1 & 10 & MDM 1.3~m  \\
turk3Octsdb & 03 Oct & 17:41:49 & 8.5 & 10 & Tubitak 1.5~m \\
mdm100405 & 04 Oct & 09:24:00 & 2.2 & 15 & MDM 1.3~m  \\
mdm100505 & 05 Oct & 02:33:00 & 9.4 & 12 & MDM 1.3~m  \\
mdm100605 & 06 Oct & 02:23:00 & 9.4 & 10 & MDM 1.3~m  \\
mdm100705 & 07 Oct & 02:15:00 & 9.6 & 12 & MDM 1.3~m  \\
lul7Oct & 07 Oct & 18:32:20 & 1.6 & 20 & Lulin 1.0~m  \\
a024 & 07 Oct & 20:28:14 & 6.2 & 10 & SAAO 1.9~m \\
mdm100805 & 08 Oct & 03:17:00 & 8.5 & 15 & MDM 1.3~m  \\
a039 & 08 Oct & 21:31:48 & 5.4 & 10 & SAAO 1.9~m \\
mdm100905 & 09 Oct & 02:17:00 & 9.3 & 10 & MDM 1.3~m  \\
a057 & 09 Oct & 20:11:53 & 4.1 & 10 & SAAO 1.9~m \\
mdm101005 & 10 Oct & 02:06:00 & 9.6 & 10 & MDM 1.3~m  \\
a077 & 10 Oct & 19:58:49 & 6.5 & 10 & SAAO 1.9~m \\
mdm101105 & 11 Oct & 02:00:00 & 9.6 & 10 & MDM 1.3~m  \\ \hline
\end{tabular}
\end{table}

\clearpage

\begin{deluxetable}{|l|c|c|c|c|}
\tablenum{5}
\tabletypesize{\footnotesize}
\tablecaption{Frequencies detected for differing subsets of PG~0048 data. 
Column 1 is the frequency designation; column 2 is the frequency determined from
a combined data set; column 3 is the formal
least-squares error from the combined data set; column 4 is the standard 
deviation of frequencies detected
from individual runs; and column 5 is the number of individual 2005
and 2004 runs in which
that frequency is detected. All frequencies are given in $\mu$Hz. 
($^{\dagger}$ indicates frequencies that were not 
detected in 2005.) \label{tab05} }
\tablehead{ \colhead{Des.} &
\colhead{Freq.} & \colhead{$\sigma_{fit}$} & \colhead{$\sigma$} & \colhead{$N$} }
\startdata
$f1$ & 5203.1 &- & 5.9 & 2  \\
$f2$ & 5244.9 & 0.16 & 4.6 & 15  \\
$f3$ & 5287.6 & 0.03 & 3.6 & 8 \\
$f4$ & 5356.9 & 0.05 & 9.8 & 8 \\
$f5$ & 5407.0 & 0.07 &  3.2 & 1 \\
$f6$ & 5465.1 &- &- & 1 \\
$f7$ & 5487.2 & 0.05 & 8.5 & 7 \\
$f8$ & 5612.2 & 0.05 & 4.5 & 8 \\
$f9$ & 5652.9 & -   &  -  &  $1^{\dagger}$ \\
$f10$ & 6609.2 &- & 9.0 & 2\\
$f11$ & 6834.3 & 0.15 & 9.0 & $3^{\dagger}$ \\
$f12$ & 7154.3 & 0.08 & 1.4 & 1  \\
$f13$ & 7237.0 & 0.04 & 7.0 & 10\\
$f14$ & 7430.1 &    -  & -  & 1 \\
$f15$ & 7501.3 & 0.07 & -  & 1 \\
$f16$ & 7523.9 & 0.06 & 15.0 & 8 \\
$f17$ & 7560.0 &   -   & -   & 1 \\
$f18$ & 7610.1 & 0.07 & 5.6 & 5 \\
$f19$ & 8055.5 & 0.06 & 10.1 & 8 \\
$f20$ & 8651.4 & 0.08 &  -  & 1 \\
$f21$ & 8820.6 & 0.08 & 6.4 & 4 \\
$f22$ & 9352.8 & 0.07 & 16.3 & 1 \\
$f23$ & 9385.3 & 0.28 & 8.9 & $1^{\dagger}$ \\
$f24$ & 9694.6 &   -  & -  & 1 \\
$f25$ & 9795.1 & 0.08 & 18.8 & 2\\
$f26$ & 10366.8 & 0.08 & 1.4 & 2  \\
$f27$ & 11103.3 &  -  &  -  &  $1^{\dagger}$ \\
$f28$ & 11159.8 & 0.07 &- & 1 \\ \hline
\enddata
\end{deluxetable}

\clearpage

\begin{deluxetable}{|l|c|c|c|c|c|c|c|c|c|c|c|c|c|c|c|c|}
\tablenum{6}
\rotate
\tabletypesize{\scriptsize}
\setlength{\tabcolsep}{0.05in}
\tablewidth{0pc}
\tablecaption{Amplitudes detected for differing subsets of PG~0048 data
during the 2005 multisite campaign. The last two rows give the $4\sigma$ 
detection limit and the temporal resolution for each subset. Note 
that all frequencies are given in $\mu$Hz. \label{tab06} Runs 1 through 12 are boao,
mdm0929, mdm9030, mdm1001, mdm1003, mdm1006, mdm1007, saao1007, mdm1008, mdm1010, 
saao1010, and mdm1011 and runs G1 through G4 are MDM only, Oct. 1 - 3, Oct. 7 - 11, and
all the 2005 data, respectively. }
\tablehead{ \colhead{Des.} &
\colhead{1} & \colhead{2} & \colhead{3} & \colhead{4} & \colhead{5} & \colhead{6} & \colhead{7} & \colhead{8} & \colhead{9} & \colhead{10} & \colhead{11} & \colhead{12} & \colhead{G1} & \colhead{G2} & \colhead{G3} & \colhead{G4}}
\startdata
%     boao m0929 m0930  m1001  m1003  m1006 m1007  s1007 m1008 m1010 s1010 m1011 MDM    Turk  saa0   2005
$f1$ &     &     &      &      &      & 1.11&      & 1.69&     &     &     &     &      &     & &  \\ \hline
$f2$ & 2.44& 1.03& 1.48 & 1.25 & 2.26 & 1.12& 0.93 & 2.34& 1.70& 2.19& 2.42& 1.77& 1.38 & 1.07& 1.77& 1.40 \\ \hline
$f3$ &     & 1.21& 1.39 & 1.22 & 1.01 & 1.26&      &     & 1.77&     &     &     & 0.60 & 0.62& 0.78& 0.83 \\ \hline
$f4$ & 1.13&     & 1.30 & 1.24 &      &     & 1.07 &     &     &     & 1.01&     & 0.59 & 0.94& 0.53& 0.54\\ \hline
$f5$ &     &     &      & 1.19 &      &     &      &     &     &     &     &     & 0.41 &     & 0.52& 0.42\\ \hline
$f6$ &     & 0.86&      &      &      &     &      &     &     &     &     &     &      &     &     & 0.59\\ \hline
$f7$ &1.06 & 1.05&      &      & 1.02 &     &      &     & 0.87& 1.59&     &     & 0.56 &     & 0.66& \\ \hline
$f8$ &     & 0.98& 1.12 & 1.39 & 1.06 &     &      &     & 1.02&     &     &     &      & 0.92&     & 0.53\\ \hline
$f10$ &    &     &      &      &      &     & 0.94 &     &     &     &     &     &      &     &     & \\ \hline
$f12$ &    & 0.81&      &      &      &     &      &     &     &     &     &     & 0.36 &     &     & \\ \hline
$f13$ &    & 1.36&      &      &      & 1.04& 0.95 & 1.68& 1.50& 1.46& 2.13&     & 1.07 & 0.79& 1.08& 0.78\\ \hline
$f14$ &    &     &      &      &      &     & 0.83 &     &     &     &     &     &      &     &     & \\ \hline
$f15$ &    &     &      &      &      &     & 1.52 &     &     &     &     &     & 0.37 & 0.64& 0.79& 0.37\\ \hline
$f16$ &    &     & 0.97 &      &      & 1.36& 2.26 &     & 1.39&     & 1.78& 1.09& 0.53 &     & 0.86& 0.48\\ \hline
$f17$ &    &     &      &      &      &     &      &     &     &     & 1.10&     & 0.50 &     &     & 0.48\\ \hline
$f18$ &    &     & 1.00 &      &      & 1.09& 1.10 &     &     &     &     & 0.99& 0.48 & 0.63& 0.59& 0.36\\ \hline
$f19$ &1.12& 0.89& 1.08 &      &      &     &      &     &     &     & 1.13& 1.74& 0.43 & 0.58& 0.69& 0.47\\ \hline
$f20$ &    &     &      &      &      &     &      &     & 0.90&     &     &     & 0.36 &     & 0.44& \\ \hline
$f21$ &    &     &      &      & 1.41 &     & 0.81 &     & 0.94&     &     &     & 0.39 & 0.63&     & \\ \hline
$f22$ &    &     &      & 1.10 &      &     &      &     &     &     &     &     & 0.41 & 0.46&     & \\ \hline
$f23$ &    &     &      &      &      &     &      &     &     &     &     &     &      &     & 0.44& \\ \hline
$f24$ &    &     &      & 0.85 &      &     &      &     &     &     &     &     &      &     &     & \\ \hline
$f25$ &    &     &      & 1.22 &      &     & 0.88 &     &     &     &     &     & 0.36 &     & 0.48& 0.35\\ \hline
$f26$ &    &     &      &      & 0.91 &     & 1.01 &     &     &     &     &     & 0.36 & 0.56&     & \\ \hline
$f28$ &    &     &      &      &      &     &      &     &     &     &     & 1.18& 0.40 &     & 0.54& 0.35\\ \hline
$4\sigma$&0.96& 0.75& 0.84 & 0.85 & 0.90 & 1.02& 0.76 & 1.60& 0.84& 1.07& 1.00& 0.91 & 0.31& 0.47& 0.43& 0.28\\ \hline
$1/T$ & 53  & 29  & 29   & 31   & 29   & 29  & 29   & 46  & 31  & 29  & 39  & 36  & 0.9 & 2.2 & 2.2 & 0.8 \\ \hline 
\enddata 
\end{deluxetable}
\clearpage

\begin{deluxetable}{|l|c|c|c|c|c|c|c|c|c|c|c|c|c|c|c|c|c|c|c|c|}
\tablenum{7}
\rotate
\tabletypesize{\scriptsize}
\setlength{\tabcolsep}{0.05in}
\tablewidth{0pc}
\tablecaption{The same as Table~\ref{tab06} for the discovery data (1997 and
1998; courtesy of C. Koen) and 2004 MDM data. \label{tab07} Runs dd1 through dd10
are tex151, tex157, tex177, tex182, tex186, dmk157, dmk163, ck261, ck263, and ck265,
runs 1 through 3 are mdr285, mdr290, and mdr295, and runs G1 through G4 are all 1997, 
October 1998, November 1998, and 2004 data, respectively. }
\tablehead{ \colhead{Des.} & \colhead{dd1}  & \colhead{dd2} & \colhead{dd3} & \colhead{dd4} & \colhead{dd5}
& \colhead{dd6} & \colhead{dd7} & \colhead{dd8}  & \colhead{dd9}  & \colhead{dd10} &
\colhead{1} & \colhead{2} & \colhead{3} & \colhead{G1} & \colhead{G2} & \colhead{G3} & \colhead{G4}}
\startdata
%     tex151 tex157 tex177 tex182tex186 dmk157 dmk163 ck261 ck263 ck265 mdr285 mdr290mdr295 1997 1998a 1998b 2004
$f1$ &      &      &      &     &     &      &      &      &     & 2.63 &     &     &     &     &     &      &   \\ \hline
$f2$ &      &      &      &     & 3.18&      &      &      &     &      & 1.11& 1.95& 2.01& 2.78&     &      & 1.65\\ \hline
$f3$ & 5.86 &      & 3.52 &     &     &      & 2.77 &      &     &      & 1.76& 4.29&     &     & 2.10&      & 1.54 \\ \hline
$f4$ &      &      & 3.39 &2.33 &     &      &      &      &     & 3.40 & 1.96& 1.14& 1.90& 1.93& 1.34&      & 1.57 \\ \hline
$f5$ &      &      &      &     &1.96 &      &      &      &     &      &     &     &     &     &     &      & 0.55 \\ \hline
$f7$ &      &      & 1.96 & 1.20& 2.12&      &      &      &     &      & 0.93& 0.84&     & 1.94&     & 2.16 & 0.75 \\ \hline
$f8$ & 4.94*& 5.80 & 3.97 & 4.39& 3.10& 3.43 & 3.57 & 3.95 & 4.96&      & 2.85& 3.10& 3.14& 4.05& 3.34& 2.01 & 2.97 \\ \hline
$f9$ &     &      &      &     &     &      &      &      &     &      &     &     &1.28 &     &     &      &  0.65\\ \hline
$f10$ &     &      &      &     &     &      &      &      &     &      &     &     & 1.18&     &     &      & 0.68 \\ \hline
$f11$ &     &      &      &     &     &      &      &      &     &      & 1.03& 1.16& 0.98&     & 1.56&      & 1.00 \\ \hline
$f13$ &     &      &      &     &     &      &      &      &     &      & 2.76& 1.55& 1.56&     &     &      & 1.56 \\ \hline
$f16$ &4.31*&      &      &     &     &      &      &      &     &      &     & 2.28& 1.22&     &     &      & 1.26 \\ \hline
$f17$ &     &      & 1.55 & 2.17&     &      &  1.69&      &     &      &     &     &     & 1.40& 1.56&      &      \\ \hline
$f18$ &     &      &      &     &     &      &      &      &     &      & 1.25&     &     &     &     &      & 0.77 \\ \hline
$f19$ &     &      &      &     & 2.36&      &      &      & 3.29&      & 1.21& 1.08&2.03 & 1.17&     & 1.99 & 1.06 \\ \hline
$f20$ &     &      &      & 1.75&     &      &      &      &     &      &     &     &     &     &     &      &      \\ \hline
$f21$ &     &      &      &     &     &      &      &      &     &      &     & 0.93&     &     &     &      & 0.74 \\ \hline
$f22$ &     &      &      & 2.43&     &      &      &      &     &      &     &     &     & 1.26&     &      &      \\ \hline
$f23$ &     &      &      &     &     &      &      &      &     &      &     & 1.04&     &     &     &      & 0.65 \\ \hline
$f24$ &     &      &      &     &     &      &      &      &     &      &     &     &     &     &     &      & 0.53\\ \hline
$f27$ &     &      &      &     &     &      &      &      &     &      &     & 1.00&     &     &     &      &      \\ \hline
$4\sigma$&5.38&3.54& 1.50 & 1.20& 1.70& 2.06 & 1.63 & 3.80 & 3.17& 2.44 & 0.75& 0.82& 0.97& 1.14& 1.21 & 1.97 & 0.50\\ \hline
$1/T$ & 194 & 132  & 69   & 96  & 84  & 96   & 84   & 174  & 185 & 111  & 41  & 38  & 37  & 0.8 & 5.4 &      & 0.9 \\ \hline
\enddata 
\end{deluxetable}
\clearpage

\begin{deluxetable}{|lll}
\tablenum{8}
\tabletypesize{\footnotesize}
\tablecaption{PG~0048 pulsation frequencies split by 
integer multiples near 972 and 41.1~$\mu$Hz.
The first column gives the minimum degree $\ell$ for the multiplet and the
order for each row is from the top of Table~\ref{tab05}, or in
descending frequencies and the number in parentheses indicates the integer
multiple between itself and the previously listed
frequency and the percent deviation from 972 or 41.1~$\mu$Hz. \label{tab08} }
%
%\begin{deluxetable}{ll}
%\tablenum{8}
%\caption{Pulsation frequencies split by a multiple near 972~$\mu$Hz.
%The order for each row is from the top of Table~\ref{tab04}, or in
%descending frequencies and the number in parentheses indicates the multiple
%of the rotation/orbital frequency between itself and the previously listed
%frequency and the percent deviation from 972~$\mu$Hz.}
%\begin{tabular}{|l|l|}
\tablehead{ \colhead{$\ell_{min}$} & \colhead{Des.} & \colhead{Designations for related frequencies} }
\startdata
 \multicolumn{3}{c}{Spacings of 972~$\mu$Hz}\\
3 & $f1$ & $f12$ (2, 0.2\%); $f27$ (4, 1.4\%) \\
3 & $f7$ & $f14$ (2, 0.2\%); $f23$ (2, 0.4\%); $f26$ (1, 0.8\%)\\
1 & $f8$ & $f10$ (1, 2.4\%); $f17$ (1, 2.4\%)\\
2 & $f9$ & $f18$(2, 0.5\%); $f20$ (1, 6.9\%) \\
2 & $f11$ & $f21$ (2, 2.0\%); $f25$ (1, 0.07\%) \\
1 & $f3$ & $f13$ (2, 0.4\%)\\
2 & $f6$ & $f22$ (4, 0.2\%)\\ \hline
\multicolumn{3}{c}{Spacings of 41.1~$\mu$Hz}\\
6 & $f1$ & $f2$ (1, 1.7\%); $f3$ (1, 3.9\%); $f5$ (3, 3.2\%); $f7$ (2, 2.4\%); $f8$ (3, 1.5\%); $f9$ (1, 0.1\%)\\
6 (1, 1) & $f12$ & $f13$ (2, 0.05\%); $f16^{\dagger}$ (7, 0.02\%); $f18$ (2, 5.9\%)\\
\hline
\enddata
\end{deluxetable}

\clearpage

\begin{deluxetable}{|lcccc|}
\tablenum{9}
\tablecaption{Results of simulations for phase changes using $f2$ as
the template and a 
comparison with observations. Column 1 indicates how the phase was
changed in the simulation and columns 2 and 3 provide the average 
amplitude and standard
deviation of $>100,000$ simulations. Columns 4 and 5 provide ratios
comparing amplitudes between individual runs and that of the combined
simulated data set.
The bottom row provides the average amplitude from the individual runs and 
results from the simulations with constant phase and fixed amplitudes and 
phases with those observed for $f2$.\label{tab09}}
\tablehead{ \colhead{Sim} & \colhead{$\langle A\rangle$} & \colhead{$\sigma$} 
& \colhead{$\langle A\rangle /\langle A\rangle_{ind}$} & 
\colhead{$\langle A\rangle /A_{max}$} }
\startdata
10\% & 1.50 & 0.03 & 0.97 & 0.66 \\
20\% & 1.35 & 0.10 & 0.88 & 0.60 \\
Random & 1.06 & 0.11 & 0.69 & 0.47 \\ \hline
\multicolumn{1}{c}{$\langle A\rangle_{ind}$} & \multicolumn{2}{c}{constant phase } & \multicolumn{2}{c}{$A_{obs}$ \& $\phi_{obs}$} \\
\multicolumn{1}{c}{1.54} & \multicolumn{2}{c}{1.55} & \multicolumn{2}{c}{1.49}\\
\enddata
\end{deluxetable}

\clearpage
\begin{deluxetable}{|l|c|c|c|c|c|c|c|c|c|c|c|c|}
\tablenum{10}
\rotate
\tabletypesize{\scriptsize}
\setlength{\tabcolsep}{0.05in}
\tablewidth{0pc}
\tablecaption{Observed properties of individual frequencies.\label{tab10} }
\tablehead{ \colhead{Des.} & \colhead{\#$_{det}$} 
& \colhead{$\langle \sigma_d\rangle$} 
& \colhead{\#$_{exp}$} &
\colhead{$\langle A\rangle$} & \colhead{$\sigma_A$} 
& \colhead{$\sigma_A/\langle A\rangle$} & \colhead{$A_{max}$} 
& \colhead{$A_{G4}/\langle A\rangle $} & \colhead{$A_{G4}/A_{max}$}
& \colhead{$A_{G1}/\langle A\rangle$ } & \colhead{$A_{G1}/A_{max}$}
& \colhead{$\sigma_{\phi}$ (\%)} }
\startdata
$f1$  & 2 & 0.38 &  3 & 1.30 & 0.41 & 0.32 & 1.69 & - & - & - & - & -\\ \hline
$f2$ & 12 & 3.69 & 12 & 1.69 & 0.53 & 0.31 & 2.44 & 0.83 & 0.57 & 0.82 & 0.57 & 9.6\\ \hline
$f3$  & 5 & 2.63 & 11 & 1.32 & 0.22 & 0.17 & 1.77 & 0.63 & 0.47 & 0.45 & 0.34 & - \\ \hline
$f4$  & 5 & 1.59 & 10 & 1.14 & 0.15 & 0.13 & 1.30 & 0.47 & 0.44 & 0.52 & 0.48 & 8.4 \\ \hline
$f5$  & 2 & 0.92 &  7 & 1.06 & 0.19 & 0.18 & 1.19 & 0.40 & 0.35 & 0.39 & 0.34 & - \\ \hline
$f6$  & 1 & 0.11 &  1 & 0.86 & - & - & 0.86 & 0.67 & 0.67 & -    & -    & - \\ \hline
$f7$  & 5 & 1.16 &  8 & 1.09 & 0.19 & 0.17  & 1.59 & -    & -    & 0.51 & 0.35 & 7.9 \\ \hline
$f8$  & 5 & 1.50 & 10 & 1.11 & 0.09 & 0.08 & 1.39 & 0.48 & 0.38 & -     & -   & 6.3 \\ \hline
$f9$  & 1 & -    & -  & -    & - & - & -    & -    & -    & -    &  -   & -  \\ \hline
$f10$ & 1 & 1.06 &  8 & 0.94 & - & - & 0.94 & -    & -    & -    &  -   & -  \\ \hline
$f11$ & 1 & -    & -  & -    & - & - & -    & -    & -    & -    &  -   & -  \\ \hline
$f12$ & 1 & 0.33 &  3 & 0.81 & - & - & 0.81 & -    & -    & 0.44 & 0.44 & - \\ \hline
$f13$ & 7 & 2.32 & 11 & 1.43 & 0.36 & 0.25 & 2.13 & 0.55 & 0.37 & 0.75 & 0.50 & 15.8 \\ \hline
$f14$ & 1 & 0.41 &  3 & 0.83 & - & - & 0.83 & -    & -    & -    &  -   & - \\ \hline
$f15$ & 1 & 2.75 & 11 & 1.52 & - & - & 1.52 & 0.24 & 0.24 & 0.24 & 0.24 & - \\ \hline
$f16$ & 6 & 2.60 & 11 & 1.38 & 0.44 & 0.32 & 2.26 & 0.35 & 0.21 & 0.38 & 0.23 & 15.0 \\ \hline
$f17$ & 1 & 0.50 &  4 & 1.10 & - & - & 1.10 & 0.44 & 0.44 & 0.45 & 0.45 & - \\ \hline
$f18$ & 4 & 0.94 &  7 & 1.05 & 0.18 & 0.17 & 1.10 & 0.43 & 0.33 & 0.46 & 0.44 & 25.9\\ \hline
$f19$ & 5 & 1.79 & 10 & 1.19 & 0.26 & 0.22 & 1.74 & 0.39 & 0.27 & 0.36 & 0.25 & 7.3 \\ \hline
$f20$ & 1 & 0.32 &  2 & 0.90 & - & - & 0.90 & -    & -    & 0.40 & 0.40 & - \\ \hline
$f21$ & 3 & 1.16 &  8 & 1.03 & 0.26 & 0.25 & 1.41 & -    & -    & 0.38 & 0.28 & - \\ \hline
$f22$ & 1 & 1.32 &  9 & 1.10 & - & - & 1.10 & -    & -    & 0.37 & 0.37 &  - \\ \hline
$f23$ & 1 & -    & -  & -    & - & - & -    & -    & -    & -    &  -   & -  \\ \hline
$f24$ & 1 & 0.00 & 1  & -    & - & - & -    & -    & -    & -    &  -   & -  \\ \hline
$f25$ & 2 & 2.25 & 11 & 1.03 & 0.24 & 0.23 & 1.22 & 0.34 & 0.29 & 0.35 & 0.30 & - \\ \hline
$f26$ & 3 & 0.76 &  6 & 0.94 & 0.17 & 0.18 & 1.01 & -    & -    & 0.41 & 0.36 & - \\ \hline
$f27$ & 1 & -    & -  & -    & - & - & -    & -    & -    & -    &  -   & -  \\ \hline
$f28$ & 1 & 1.42 &  9 & 1.18 & - & - & 1.18 & 0.30 & 0.30 & 0.34 & 0.34 & - \\ \hline
\enddata
\end{deluxetable}

\clearpage
%% the file, uncomment the \plotone, \plottwo, and

\clearpage

\clearpage

\clearpage

\clearpage

\clearpage

\clearpage

\clearpage

\clearpage

\clearpage

\clearpage

 \clearpage 
% 
%\clearpage 
%\begin{figure}
%\figurenum{12}
%\includegraphics[angle=-90,width=\textwidth]{3split.ps}
%\caption{Temporal spectra in $\sim 4$ hour
%segments from the beginning (top), middle (middle) and end (bottom) of 
%individual PG~0048 observing runs. Horizontal lines indicate the $4\sigma$
%detection limit.}
%\label{fig12}
%\end{figure}

\clearpage

\clearpage

\clearpage
\figurenum{15}
\includegraphics[angle=-90,width=\textwidth]{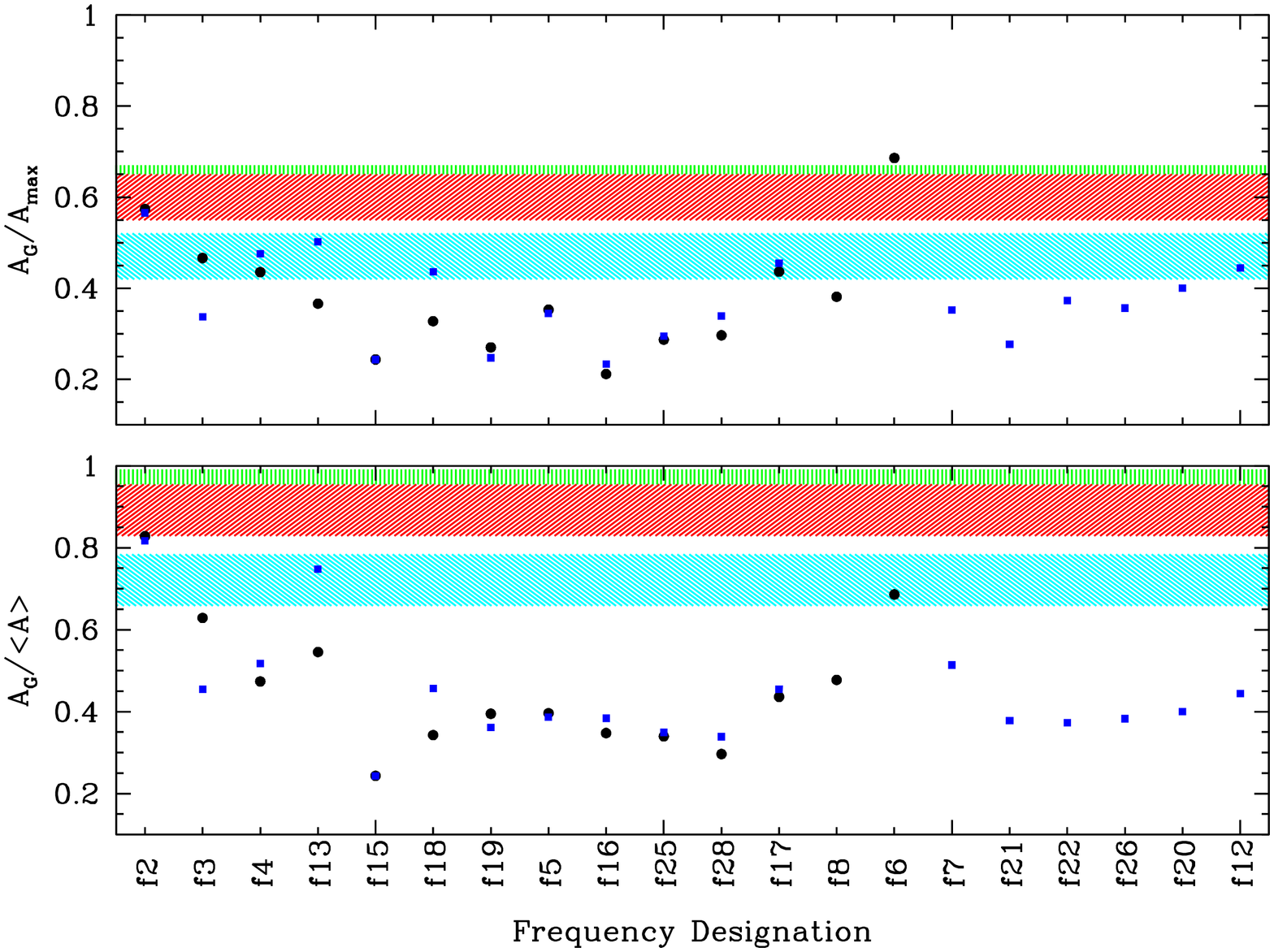}
%\plottwo{f14aa.ps}{f14b.ps}
\figcaption{Comparison of amplitude ratios from simulations to those observed
for PG~0048. Black circles are for G4 and blue squares are for G1 data.
The verticle-lined (top and green) area is for simulations 
with 10\% phase variations, the $+45^o$-lined (middle and red) area 
is for simulations with 20\% phase variations, and
the $-45^o$-lined (bottom and blue) area is for simulations with random phases.
\label{fig15}}

\clearpage

\clearpage
\figurenum{17}
\includegraphics[angle=-90,width=\textwidth]{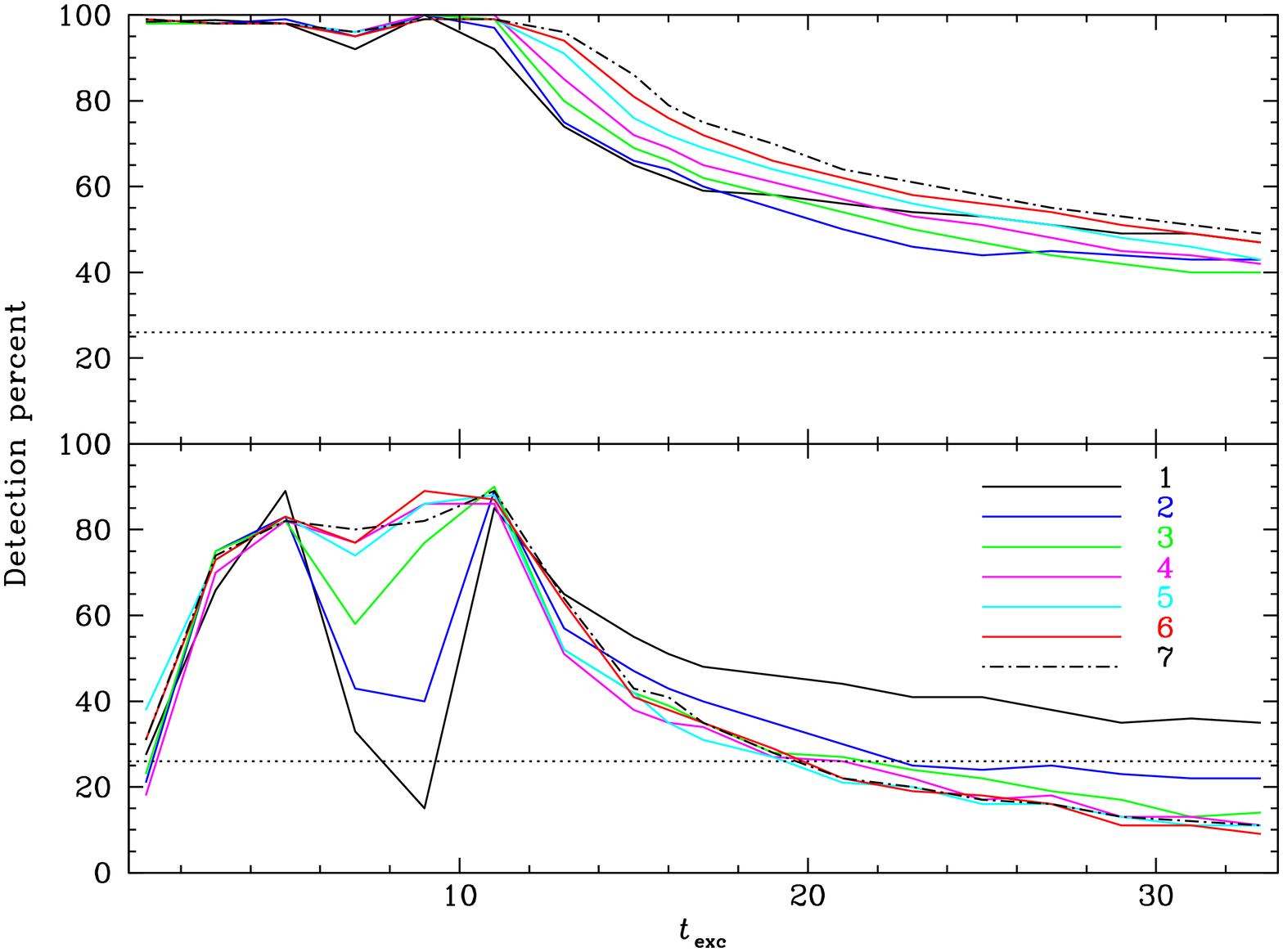}
\figcaption{Expected fraction of detected frequencies based on simulations. Top
panel used average amplitudes while the bottom panel used maximum
amplitudes from the simulations. Differing lines represent different values of
$t_d$ given in the legend. Dotted line is the observed 26\% detection rate.
See the electronic edition of the Journal for a color version 
of this figure.
\label{fig17}}

\end{document}